\documentclass[11pt,a4paper]{article}
\pdfoutput=1

\usepackage{jheppub}

\usepackage[utf8]{inputenc}
\usepackage[T1]{fontenc}
\usepackage{afterpage}

\newcommand{\Dmq}{\Delta m^2}
\newcommand{\Eps}{\varepsilon}
\newcommand{\Epx}{\mathcal{E}}
\newcommand{\diag}{\mathop{\mathrm{diag}}}
\renewcommand{\Re}{\mathop{\mathrm{Re}}}

\newenvironment{pagefigure}{\begin{figure}[!p]}{\afterpage\clearpage\end{figure}}

\title{Updated Constraints on Non-Standard Interactions from Global
  Analysis of Oscillation Data}

\author[a]{Ivan Esteban,}
\affiliation[a] {Departament de Fis\'{\i}ca Qu\`antica i
  Astrof\'{\i}sica and Institut de Ciencies del Cosmos, Universitat de
  Barcelona, Diagonal 647, E-08028 Barcelona, Spain}
\emailAdd{ivan.esteban@fqa.ub.edu}

\author[a,b,c]{M.~C.~Gonzalez-Garcia,}
\affiliation[b]{Instituci\'o Catalana de Recerca i Estudis
  Avan\c{c}ats (ICREA), Pg. Lluis Companys 23, 08010 Barcelona,
  Spain}
\affiliation[c]{C.N.~Yang Institute for Theoretical Physics, State
  University of New York at Stony Brook, Stony Brook, NY 11794-3840,
  USA}
\emailAdd{maria.gonzalez-garcia@stonybrook.edu}

\author[d]{Michele Maltoni,}
\affiliation[d]{Instituto de F\'{\i}sica Te\'orica UAM/CSIC, Calle de
  Nicol\'as Cabrera 13--15, Universidad Aut\'onoma de Madrid,
  Cantoblanco, E-28049 Madrid, Spain}
\emailAdd{michele.maltoni@csic.es}

\author[d]{Ivan Martinez-Soler,}
\emailAdd{ivanj.m@csic.es}

\author[a]{Jordi Salvado}
\emailAdd{jor.salvado@gmail.com}

\abstract{We quantify our present knowledge of the size and flavor
  structure of non-standard neutrino interactions which affect the
  matter background in the evolution of solar, atmospheric, reactor
  and long-baseline accelerator neutrinos as determined by a global
  analysis of oscillation data --- both alone and in combination with
  the results on coherent neutrino--nucleus scattering from the
  COHERENT experiment. We consider general neutral current neutrino
  interactions with quarks whose lepton-flavor structure is
  independent of the quark type. We study the dependence of the
  allowed ranges of non-standard interaction coefficients, the status
  of the LMA-D solution, and the determination of the oscillation
  parameters on the relative strength of the non-standard couplings to
  up and down quarks. Generically we find that the conclusions are
  robust for a broad spectrum of up-to-down strengths, and we identify
  and quantify the exceptional cases related to couplings whose effect
  in neutrino propagation in the Earth or in the Sun is severely
  suppressed. As a result of the study we provide explicit constraints
  on the effective couplings which parametrize the non-standard Earth
  matter potential relevant for long-baseline experiments.}


\preprint{IFT-UAM/CSIC-18-049, YITP-SB-18-11}

\arxivnumber{1805.04530}

\begin{document}

\maketitle

\section{Introduction}

Experiments measuring the flavor composition of neutrinos produced in
the Sun, in the Earth's atmosphere, in nuclear reactors and in
particle accelerators have established that lepton flavor is not
conserved in neutrino propagation, but it oscillates with a wavelength
which depends on distance and energy. This demonstrates beyond doubt
that neutrinos are massive and that the mass states are non-trivial
admixtures of flavor states~\cite{Pontecorvo:1967fh, Gribov:1968kq},
see Ref.~\cite{GonzalezGarcia:2007ib} for an overview.

Under the assumption that the Standard Model (SM) is the low energy
effective model of a complete high energy theory, neutrino masses
emerge naturally as the first observable consequence in the form of
the Weinberg operator~\cite{Weinberg:1979sa}, the only dimension five
operator that can be built within the SM particle content.  In this
framework the next operators with observable consequences at low
energies come at dimension six.  They include four-fermion terms
leading to Non-Standard Interactions (NSI)~\cite{Wolfenstein:1977ue,
  Valle:1987gv, Guzzo:1991hi} between neutrinos and matter (for recent
reviews, see~\cite{Ohlsson:2012kf, Miranda:2015dra}), both in
charge-current interactions (NSI-CC)
\begin{equation}
  \label{eq:NSI-cc}
  (\bar\nu_\alpha \gamma_\mu P_L \ell_\beta) (\bar f' \gamma^\mu P f)
\end{equation}
and in neutral current interactions (NSI-NC)
\begin{equation}
  \label{eq:NSI-nc}
  (\bar\nu_\alpha \gamma_\mu P_L \nu_\beta) (\bar f \gamma^\mu P f) \,.
\end{equation}
Here $\alpha,\beta$ are lepton flavor indices, $f,f'$ are SM charged
fermions and $\gamma^\mu$ are the Dirac gamma matrices; $P_L$ is the
left-handed projection operator while $P$ can be either $P_L$ or $P_R$
(the right-handed projection operator). These operators are expected
to arise generically from the exchange of some mediator state assumed
to be heavier that the characteristic momentum transfer in the $\nu$
interaction process.

Since operators in both Eqs.~\eqref{eq:NSI-cc} and~\eqref{eq:NSI-nc}
modify the inelastic neutrino scattering cross sections with other SM
fermions they can be bounded by precision electroweak data (see for
example Refs.~\cite{Davidson:2003ha, Biggio:2009nt, Biggio:2009kv}).
In general these ``scattering'' bounds on NSI-CC operators are rather
stringent, whereas the bounds on NSI-NC tend to be weaker.
On the other hand, the operators in Eq.~\eqref{eq:NSI-nc} can also
modify the forward-coherent scattering (\textit{i.e.}, at zero
momentum transfer) of neutrinos as they propagate through matter via
so-called Mikheev-Smirnov-Wolfenstein (MSW)
mechanism~\cite{Wolfenstein:1977ue, Mikheev:1986gs}. Consequently
their effect can be significantly enhanced in oscillation experiments
where neutrinos travel large regions of matter, such as is the case
for solar and atmospheric neutrinos. Indeed, the global analysis of
data from oscillation experiments in the framework of mass induced
oscillations in presence of NSI currently provides some of the
strongest constraints on the size of the NSI affecting neutrino
propagation~\cite{GonzalezGarcia:2011my, Gonzalez-Garcia:2013usa}.

Of course, for models with a high energy New Physics scale,
electroweak gauge invariance generically implies that the NSI-NC
parameters are still expected to be subject to tight constraints from
charged lepton observables~\cite{Gavela:2008ra, Antusch:2008tz},
leading to no visible effect in oscillations. However, more recently
it has been argued that viable gauge models with light mediators
(\textit{i.e.}, below the electroweak scale) may lead to observable
effects in oscillations without entering in conflict with other
bounds~\cite{Farzan:2015doa, Farzan:2015hkd, Babu:2017olk,
  Farzan:2017xzy, Denton:2018xmq} (see also
Ref.~\cite{Miranda:2015dra} for a discussion). In particular, for
light mediators bounds from high-energy neutrino scattering
experiments such as CHARM~\cite{Dorenbosch:1986tb} and
NuTeV~\cite{Zeller:2001hh} do not apply.  In this framework NSI-NC
generated by mediators as light as about 10~MeV can only be
constrained by their effect in oscillation data and by the recent
results on coherent neutrino--nucleus scattering observed for the
first time by the COHERENT experiment~\cite{Akimov:2017ade}.

In this work we revisit our current knowledge of the size and flavor
structure of NSI-NC which affect the matter background in the
evolution of solar, atmospheric, reactor and long-baseline (LBL)
accelerator neutrinos as determined by a global analysis of
oscillation data. This updates and extends the analysis in
Ref.~\cite{Gonzalez-Garcia:2013usa} where NSI-NC with either up or
down quarks were considered. Here we extend our previous study to
account for the possibility of NSI with up \emph{and} down quarks
simultaneously, under the simplifying assumption that they carry the
same lepton flavor structure. To this aim, in Sec.~\ref{sec:formalism}
we briefly summarize the framework of our study and discuss the
simplifications used in the analysis of the atmospheric and LBL data
on one side and of the solar and KamLAND sector on the other side. In
Sec.~\ref{sec:solar} we present the results of the updated analysis of
solar and KamLAND data and quantify the impact of the modified matter
potential on the data description, as well as the status of the LMA-D
solution~\cite{Miranda:2004nb} in presence of the most general NSI
scenario considered here.  In Sec.~\ref{sec:globalosc} we describe the
constraints implied by the analysis of atmospheric, LBL and reactor
experiments, and combine them with those arising from the
solar+KamLAND data.  We show how the complementarity and synergy of
the different data sets is important for a robust determination of
neutrino masses and mixing in the presence of these general NSI, and
we derive the most up-to-date allowed ranges on NSI couplings. Finally
in Sec.~\ref{sec:coherent} we further combine the oscillation bounds
with those from the COHERENT experiment and in Sec.~\ref{sec:summary}
we summarize our conclusions. We present the details of the analysis
of the IceCube results in Appendix~\ref{sec:icecube}.

\section{Formalism}
\label{sec:formalism}

In this work we will consider NSI affecting neutral-current processes
relevant to neutrino propagation in matter. The coefficients
accompanying the new operators are usually parametrized in the form:
\begin{equation}
  \label{eq:NSILagrangian}
  \mathcal L_\text{NSI} = -2\sqrt2 G_F
  \sum_{f,P,\alpha,\beta} \Eps_{\alpha\beta}^{f,P}
  (\bar\nu_\alpha\gamma^\mu P_L\nu_\beta)
  (\bar f\gamma_\mu P f) \,,
\end{equation}
where $G_F$ is the Fermi constant, $\alpha, \beta$ are flavor indices,
$P\equiv P_L, P_R$ and $f$ is a SM charged fermion. In this notation,
$\Eps_{\alpha\beta}^{f,P}$ parametrizes the strength of the new
interaction with respect to the Fermi constant,
$\Eps_{\alpha\beta}^{f,P} \sim \mathcal{O}(G_X/G_F)$.
If we now assume that the neutrino flavor structure of the
interactions is independent of the charged fermion type, we can
factorize $\Eps_{\alpha\beta}^{f,P}$ as the product of two terms:
\begin{equation}
  \label{eq:eps-fact}
  \Eps_{\alpha\beta}^{f,P} \equiv \Eps_{\alpha\beta}^\eta \, \xi^{f,P}
\end{equation}
where the matrix $\Eps_{\alpha\beta}^\eta$ describes the neutrino part
and the coefficients $\xi^{f,P}$ parametrize the coupling to the
charged fermions. Under this assumption the Lagrangian in
Eq.~\eqref{eq:NSILagrangian} takes the form:
\begin{equation}
  \mathcal L_\text{NSI} = -2\sqrt2 G_F
  \bigg[ \sum_{\alpha,\beta} \Eps_{\alpha\beta}^\eta
  (\bar\nu_\alpha\gamma^\mu P_L\nu_\beta) \bigg]
  \bigg[ \sum_{f,P} \xi^{f,P} (\bar f\gamma_\mu P f) \bigg] \,.
\end{equation}
As is well known, only vector NSI contribute to the matter potential
in neutrino oscillations. It is therefore convenient to define:
\begin{equation}
  \label{eq:eps-xi}
  \Eps_{\alpha\beta}^f
  \equiv \Eps_{\alpha\beta}^{f,L} + \Eps_{\alpha\beta}^{f,R} 
  = \Eps_{\alpha\beta}^\eta \, \xi^f
  \quad\text{with}\quad
  \xi^f \equiv \xi^{f,L} + \xi^{f,R} \,.
\end{equation}
Ordinary matter is composed of electrons ($e$), up quarks ($u$) and
down quarks ($d$). As stated in the introduction, in this work we
restrict ourselves to non-standard interactions with quarks, so that
only $\xi^u$ and $\xi^d$ are relevant for neutrino propagation. It is
clear that a global rescaling of both $\xi^u$ and $\xi^d$ by a common
factor can be reabsorbed into a rescaling of
$\Eps_{\alpha\beta}^\eta$, so that only the direction in the $(\xi^u,
\xi^d)$ plane is phenomenologically non-trivial. We parametrize such
direction in terms of an angle $\eta$, which for later convenience we
have related to the NSI couplings of protons and neutrons described in
the next section (see Eqs.~\eqref{eq:eps-nucleon}
and~\eqref{eq:epx-eta} for a formal definition). In terms of the
``quark'' couplings introduced in Eq.~\eqref{eq:eps-xi} we have:
\begin{equation}
  \label{eq:xi-eta}
  \xi^u = \frac{\sqrt{5}}{3} (2 \cos\eta - \sin\eta) \,,
  \qquad
  \xi^d = \frac{\sqrt{5}}{3} (2 \sin\eta - \cos\eta)
\end{equation}
where we have chosen the normalization so that $\eta = \arctan(1/2)
\approx 26.6^\circ$ corresponds to NSI with up quarks ($\xi^u=1$,
$\xi^d=0$) while $\eta = \arctan(2) \approx 63.4^\circ$ corresponds to
NSI with down quarks ($\xi^u=0$, $\xi^d=1$). Note that the
transformation $\eta \to \eta + \pi$ simply results in a sign flip of
$\xi^u$ and $\xi^d$, hence it is sufficient to consider $-\pi/2 \leq
\eta \leq \pi/2$.

\subsection{Neutrino oscillations in the presence of NSI}

In general, the evolution of the neutrino and antineutrino flavor
state during propagation is governed by the Hamiltonian:
\begin{equation}
  H^\nu = H_\text{vac} + H_\text{mat}
  \quad\text{and}\quad
  H^{\bar\nu} = ( H_\text{vac} - H_\text{mat} )^* \,,
\end{equation}
where $H_\text{vac}$ is the vacuum part which in the flavor basis
$(\nu_e, \nu_\mu, \nu_\tau)$ reads
\begin{equation}
  \label{eq:Hvac}
  H_\text{vac} = U_\text{vac} D_\text{vac} U_\text{vac}^\dagger
  \quad\text{with}\quad
  D_\text{vac} = \frac{1}{2E_\nu} \diag(0, \Dmq_{21}, \Dmq_{31}) \,.
\end{equation}
Here $U_\text{vac}$ denotes the three-lepton mixing matrix in
vacuum~\cite{Pontecorvo:1967fh, Maki:1962mu, Kobayashi:1973fv}.
Following the convention of Ref.~\cite{Coloma:2016gei}, we define
$U_\text{vac} = R_{23}(\theta_{23}) R_{13}(\theta_{13})
\tilde{R}_{12}(\theta_{12}, \delta_\text{CP})$, where
$R_{ij}(\theta_{ij})$ is a rotation of angle $\theta_{ij}$ in the $ij$
plane and $\tilde{R}_{12}(\theta_{12},\delta_\text{CP})$ is a complex
rotation by angle $\theta_{12}$ and phase
$\delta_\text{CP}$. Explicitly:
\begin{equation}
  \label{eq:Uvac}
  U_\text{vac} =
  \begin{pmatrix}
    c_{12} c_{13}
    & s_{12} c_{13} e^{i\delta_\text{CP}}
    & s_{13}
    \\
    - s_{12} c_{23} e^{-i\delta_\text{CP}} - c_{12} s_{13} s_{23}
    & \hphantom{+} c_{12} c_{23} - s_{12} s_{13} s_{23} e^{i\delta_\text{CP}}
    & c_{13} s_{23}
    \\
    \hphantom{+} s_{12} s_{23} e^{-i\delta_\text{CP}} - c_{12} s_{13} c_{23}
    & - c_{12} s_{23} - s_{12} s_{13} c_{23} e^{i\delta_\text{CP}}
    & c_{13} c_{23}
  \end{pmatrix}
\end{equation}
where $c_{ij} \equiv \cos\theta_{ij}$ and $s_{ij} \equiv
\sin\theta_{ij}$. This expression differs from the usual one ``$U$''
(defined, \textit{e.g.}, in Eq.~(1.1) of Ref.~\cite{Esteban:2016qun})
by an overall phase matrix:
\begin{equation}
  U_\text{vac} = P U P^*
  \quad\text{with}\quad
  P = \diag(e^{i\delta_\text{CP}}, 1, 1) \,.
\end{equation}
It is easy to show that in the absence of non-standard interactions
such rephasing does not affect the expression of the probabilities and
produces therefore no visible effect: in other words, when only
standard interactions are considered the physical interpretation of
the vacuum parameters ($\Dmq_{21}$, $\Dmq_{31}$, $\theta_{12}$,
$\theta_{13}$, $\theta_{23}$ and $\delta_\text{CP}$) is exactly the
same in both conventions.
The advantage of defining $U_\text{vac}$ as in Eq.~\eqref{eq:Uvac} is
that the CPT transformation $H_\text{vac} \to -H_\text{vac}^*$, whose
relevance for the present work will be discussed below, can be
implemented exactly (up to an irrelevant multiple of the identity) by
the following transformation of the parameters:
\begin{equation}
  \label{eq:osc-deg}
  \begin{aligned}
    \Dmq_{31} &\to -\Dmq_{31} + \Dmq_{21} = -\Dmq_{32} \,,
    \\
    \theta_{12} & \to \pi/2 - \theta_{12} \,,
    \\
    \delta_\text{CP} &\to \pi - \delta_\text{CP}
  \end{aligned}
\end{equation}
which does not spoil the commonly assumed restrictions on the range of
the vacuum parameters ($\Dmq_{21} > 0$ and $0 \leq \theta_{ij} \leq
\pi/2$).

Concerning the matter part $H_\text{mat}$ of the Hamiltonian which
governs neutrino oscillations, if all possible operators in
Eq.~\eqref{eq:NSILagrangian} are added to the SM Lagrangian we get:
\begin{equation}
  \label{eq:Hmat}
  H_\text{mat} = \sqrt{2} G_F N_e(x)
  \begin{pmatrix}
    1+\Epx_{ee}(x) & \Epx_{e\mu}(x) & \Epx_{e\tau}(x) \\
    \Epx_{e\mu}^*(x) & \Epx_{\mu\mu}(x) & \Epx_{\mu\tau}(x) \\
    \Epx_{e\tau}^*(x) & \Epx_{\mu\tau}^*(x) & \Epx_{\tau\tau}(x)
  \end{pmatrix}
\end{equation}
where the ``$+1$'' term in the $ee$ entry accounts for the standard
contribution, and
\begin{equation}
  \label{eq:epx-nsi}
  \Epx_{\alpha\beta}(x) = \sum_{f=e,u,d}
  \frac{N_f(x)}{N_e(x)} \Eps_{\alpha\beta}^f
\end{equation}
describes the non-standard part. Here $N_f(x)$ is the number density
of fermion $f$ as a function of the distance traveled by the neutrino
along its trajectory.  In Eq.~\eqref{eq:epx-nsi} we have limited the
sum to the charged fermions present in ordinary matter,
$f=e,u,d$. Since quarks are always confined inside protons ($p$) and
neutrons ($n$), it is convenient to define:
\begin{equation}
  \label{eq:eps-nucleon}
  \Eps_{\alpha\beta}^p = 2\Eps_{\alpha\beta}^u + \Eps_{\alpha\beta}^d \,,
  \qquad
  \Eps_{\alpha\beta}^n = 2\Eps_{\alpha\beta}^d + \Eps_{\alpha\beta}^u \,.
\end{equation}
Taking into account that $N_u(x) = 2N_p(x) + N_n(x)$ and $N_d(x) =
N_p(x) + 2N_n(x)$, and also that matter neutrality implies $N_p(x) =
N_e(x)$, Eq.~\eqref{eq:epx-nsi} becomes:
\begin{equation}
  \label{eq:epx-nuc}
  \Epx_{\alpha\beta}(x) =
  \big( \Eps_{\alpha\beta}^e + \Eps_{\alpha\beta}^p \big)
  + Y_n(x) \Eps_{\alpha\beta}^n
  \quad\text{with}\quad
  Y_n(x) \equiv \frac{N_n(x)}{N_e(x)}
\end{equation}
which shows that from the phenomenological point of view the
propagation effects of NSI with electrons can be mimicked by NSI with
quarks by means of a suitable combination of up-quark and down-quark
contributions. Our choice of neglecting $\Eps_{\alpha\beta}^e$ in this
work does not therefore imply a loss of generality.

Since this matter term can be determined by oscillation experiments
only up to an overall multiple of the identity, each
$\Eps_{\alpha\beta}^f$ matrix introduces 8 new parameters: two
differences of the three diagonal real parameters (\textit{e.g.},
$\Eps_{ee}^f - \Eps_{\mu\mu}^f$ and $\Eps_{\tau\tau}^f -
\Eps_{\mu\mu}^f$) and three off-diagonal complex parameters
(\textit{i.e.}, three additional moduli and three complex phases).
Under the assumption that the neutrino flavor structure of the
interactions is independent of the charged fermion type, as described
in Eq.~\eqref{eq:eps-fact}, we can write $\Eps_{\alpha\beta}^p =
\Eps_{\alpha\beta}^\eta \, \xi^p$ and $\Eps_{\alpha\beta}^n =
\Eps_{\alpha\beta}^\eta \, \xi^n$, which leads to:
\begin{equation}
  \label{eq:epx-eta}
  \Epx_{\alpha\beta}(x) =
  \Eps_{\alpha\beta}^\eta \big[ \xi^p + Y_n(x) \xi^n \big]
  \quad\text{with}\quad
  \xi^p = \sqrt{5} \cos\eta
  \quad\text{and}\quad
  \xi^n = \sqrt{5} \sin\eta
\end{equation}
so that the phenomenological framework adopted here is characterized
by 9 matter parameters: eight related to the matrix
$\Eps_{\alpha\beta}^\eta$ plus the direction $\eta$ in the
$(\xi^p,\xi^n)$ plane.

We finish this section by reminding that as a consequence of the CPT
symmetry, neutrino evolution is invariant if the Hamiltonian $H^\nu =
H_\text{vac} + H_\text{mat}$ is transformed as $H^\nu \to -(H^\nu)^*$.
This requires a simultaneous transformation of both the vacuum and the
matter terms. The transformation of $H_\text{vac}$ is described in
Eq.~\eqref{eq:osc-deg} and involves a change in the octant of
$\theta_{12}$ as well as a change in the neutrino mass ordering
(\textit{i.e.}, the sign of $\Dmq_{31}$), which is why it has been
called ``generalized mass ordering degeneracy'' in
Ref.~\cite{Coloma:2016gei}. As for $H_\text{mat}$ we need:
\begin{equation}
  \label{eq:NSI-deg}
  \begin{aligned}
    \big[ \Epx_{ee}(x) - \Epx_{\mu\mu}(x) \big]
    &\to - \big[ \Epx_{ee}(x) - \Epx_{\mu\mu}(x) \big] - 2  \,,
    \\
    \big[ \Epx_{\tau\tau}(x) - \Epx_{\mu\mu}(x) \big]
    &\to -\big[ \Epx_{\tau\tau}(x) - \Epx_{\mu\mu}(x) \big] \,,
    \\
    \Epx_{\alpha\beta}(x)
    &\to - \Epx_{\alpha\beta}^*(x) \qquad (\alpha \neq \beta) \,,
  \end{aligned}
\end{equation}
see Refs.~\cite{Gonzalez-Garcia:2013usa, Bakhti:2014pva,
  Coloma:2016gei}. As seen in Eqs.~\eqref{eq:epx-nsi},
\eqref{eq:epx-nuc} and~\eqref{eq:epx-eta} the matrix
$\Epx_{\alpha\beta}(x)$ depends on the chemical composition of the
medium, which may vary along the neutrino trajectory, so that in
general the condition in Eq.~\eqref{eq:NSI-deg} is fulfilled only in
an approximate way. The degeneracy becomes exact in the following two
cases:\footnote{Strictly speaking, Eq.~\eqref{eq:NSI-deg} can be
  satisfied exactly for \emph{any} matter chemical profile $Y_n(x)$ if
  $\Eps_{\alpha\beta}^u$ and $\Eps_{\alpha\beta}^d$ are allowed to
  transform independently of each other. This possibility, however, is
  incompatible with the factorization constraint of
  Eq.~\eqref{eq:eps-fact}, so it will not be discussed here.}
\begin{itemize}
\item if the effective NSI coupling to neutrons vanishes, so that
  $\Eps_{\alpha\beta}^n = 0$ in Eq.~\eqref{eq:epx-nuc}. In terms of
  fundamental quantities this occurs when $\Eps_{\alpha\beta}^u = -2
  \Eps_{\alpha\beta}^d$, \textit{i.e.}, the NSI couplings are
  proportional to the electric charge of quarks.  In our
  parametrization this corresponds to $\eta=0$ as shown in
  Eqs.~\eqref{eq:xi-eta} and~\eqref{eq:epx-eta};

\item if the neutron/proton ratio $Y_n(x)$ is constant along the
  entire neutrino propagation path. This is certainly the case for
  reactor and long-baseline experiments, where only the Earth's mantle
  is involved, and to a good approximation also for atmospheric
  neutrinos, since the differences in chemical composition between
  mantle and core can safely be neglected in the context of
  NSI~\cite{GonzalezGarcia:2011my}. In this case the matrix
  $\Epx_{\alpha\beta}(x)$ becomes independent of $x$ and can be
  regarded as a new phenomenological parameter, as we will describe in
  Sec.~\ref{sec:formalism-earth}.
\end{itemize}
Further details on the implications of this degeneracy for different
classes of neutrino experiments (solar, atmospheric, \textit{etc.})
will be provided later in the corresponding section.

\subsection{Matter potential in atmospheric and long-baseline neutrinos}
\label{sec:formalism-earth}
  
As discussed in Ref.~\cite{GonzalezGarcia:2011my}, in the Earth the
neutron/proton ratio $Y_n(x)$ which characterize the matter chemical
composition can be taken to be constant to very good approximation.
The PREM model~\cite{Dziewonski:1981xy} fixes $Y_n = 1.012$ in the
Mantle and $Y_n = 1.137$ in the Core, with an average value
$Y_n^\oplus = 1.051$ all over the Earth. Setting therefore $Y_n(x)
\equiv Y_n^\oplus$ in Eqs.~\eqref{eq:epx-nsi} and~\eqref{eq:epx-nuc}
we get $\Epx_{\alpha\beta}(x) \equiv \Eps_{\alpha\beta}^\oplus$ with:
\begin{equation}
  \Eps_{\alpha\beta}^\oplus
  = \Eps_{\alpha\beta}^e + \big( 2 + Y_n^\oplus \big) \Eps_{\alpha\beta}^u
  + \big( 1 + 2Y_n^\oplus \big) \Eps_{\alpha\beta}^d
  = \big( \Eps_{\alpha\beta}^e + \Eps_{\alpha\beta}^p \big)
  + Y_n^\oplus \Eps_{\alpha\beta}^n \,.
\end{equation}
If we drop $\Eps_{\alpha\beta}^e$ and impose quark-lepton
factorization as in Eq.~\eqref{eq:epx-eta} we get:
\begin{equation}
  \label{eq:eps-earth}
  \Eps_{\alpha\beta}^\oplus
  = \Eps_{\alpha\beta}^\eta \big( \xi^p + Y_n^\oplus \xi^n \big)
  = \sqrt{5} \left( \cos\eta + Y_n^\oplus \sin\eta \right)
  \Eps_{\alpha\beta}^\eta \,.
\end{equation}
In other words, within this approximation the analysis of atmospheric
and LBL neutrinos holds for any combination of NSI with up, down or
electrons and it can be performed in terms of the effective NSI
couplings $\Eps_{\alpha\beta}^\oplus$, which play the role of
phenomenological parameters. In particular, the best-fit value and
allowed ranges of $\Eps_{\alpha\beta}^\oplus$ are independent of
$\eta$, while the bounds on the physical quantities
$\Eps_{\alpha\beta}^\eta$ simply scale as $(\cos\eta + Y_n^\oplus
\sin\eta)$. Moreover, it is immediate to see that for $\eta =
\arctan(-1/Y_n^\oplus) \approx -43.6^\circ$ the contribution of NSI to
the matter potential vanishes, so that no bound on
$\Eps_{\alpha\beta}^\eta$ can be derived from atmospheric and LBL data
in such case.

Following the approach of Ref.~\cite{GonzalezGarcia:2011my}, the
matter Hamiltonian $H_\text{mat}$, given in Eq.~\eqref{eq:Hmat} after
setting $\Epx_{\alpha\beta}(x) \equiv \Eps_{\alpha\beta}^\oplus$, can
be parametrized in a way that mimics the structure of the vacuum
term~\eqref{eq:Hvac}:
\begin{equation}
  \label{eq:HmatGen}
  H_\text{mat} = Q_\text{rel} U_\text{mat} D_\text{mat}
  U_\text{mat}^\dagger Q_\text{rel}^\dagger
  \text{~~with~~}
  \left\lbrace
  \begin{aligned}
    Q_\text{rel} &= \diag\left(
    e^{i\alpha_1}, e^{i\alpha_2}, e^{-i\alpha_1 -i\alpha_2} \right),
    \\
    U_\text{mat} &= R_{12}(\varphi_{12}) R_{13}(\varphi_{13})
    \tilde{R}_{23}(\varphi_{23}, \delta_\text{NS}) \,,
    \\
    D_\text{mat} &= \sqrt{2} G_F N_e(x) \diag(\Eps_\oplus, \Eps_\oplus', 0)
  \end{aligned}\right.
\end{equation}
where $R_{ij}(\varphi_{ij})$ is a rotation of angle $\varphi_{ij}$ in
the $ij$ plane and $\tilde{R}_{23}(\varphi_{23},\delta_\text{NS})$ is
a complex rotation by angle $\varphi_{23}$ and phase
$\delta_\text{NS}$.
Note that the two phases $\alpha_1$ and $\alpha_2$ included in
$Q_\text{rel}$ are not a feature of neutrino-matter interactions, but
rather a relative feature of the vacuum and matter terms.
In order to simplify the analysis we neglect $\Dmq_{21}$ and also
impose that two eigenvalues of $H_\text{mat}$ are equal
($\Eps_\oplus'=0$).  The latter assumption is justified since, as
shown in Ref.~\cite{Friedland:2004ah}, strong cancellations in the
oscillation of atmospheric neutrinos occur when two eigenvalues of
$H_\text{mat}$ are equal, and it is precisely in this situation that
the weakest constraints can be placed.
Setting $\Dmq_{21} \to 0$ implies that the $\theta_{12}$ angle and the
$\delta_\text{CP}$ phase disappear from the expressions of the
oscillation probabilities, and the same happens to the $\varphi_{23}$
angle and the $\delta_\text{NS}$ phase in the limit $\Eps_\oplus' \to
0$.
Under these approximations the effective NSI couplings
$\Eps_{\alpha\beta}^\oplus$ can be parametrized as:
\begin{equation}
  \label{eq:eps_atm}
  \begin{aligned}
    \Eps_{ee}^\oplus - \Eps_{\mu\mu}^\oplus
    &= \hphantom{-} \Eps_\oplus \, (\cos^2\varphi_{12} - \sin^2\varphi_{12})
    \cos^2\varphi_{13} - 1\,,
    \\
    \Eps_{\tau\tau}^\oplus - \Eps_{\mu\mu}^\oplus
    &= \hphantom{-} \Eps_\oplus \, (\sin^2\varphi_{13}
    - \sin^2\varphi_{12} \, \cos^2\varphi_{13}) \,,
    \\
    \Eps_{e\mu}^\oplus
    &= -\Eps_\oplus \, \cos\varphi_{12} \, \sin\varphi_{12} \,
    \cos^2\varphi_{13} \, e^{i(\alpha_1 - \alpha_2)} \,,
    \\
    \Eps_{e\tau}^\oplus
    &= -\Eps_\oplus \, \cos\varphi_{12} \, \cos\varphi_{13} \,
    \sin\varphi_{13} \, e^{i(2\alpha_1 + \alpha_2)} \,,
    \\
    \Eps_{\mu\tau}^\oplus
    &= \hphantom{-} \Eps_\oplus \, \sin\varphi_{12} \, \cos\varphi_{13} \,
    \sin\varphi_{13} \, e^{i(\alpha_1 + 2\alpha_2)} \,.
  \end{aligned}
\end{equation}
With all this the relevant flavor transition probabilities for
atmospheric and LBL experiments depend on eight parameters:
($\Dmq_{31}$, $\theta_{13}$, $\theta_{23}$) for the vacuum part,
($\Eps_\oplus$, $\varphi_{12}$, $\varphi_{13}$) for the matter part,
and ($\alpha_1$, $\alpha_2$) as relative phases.  Notice that in this
case only the relative sign of $\Dmq_{31}$ and $\Eps_\oplus$ is
relevant for atmospheric and LBL neutrino oscillations: this is just a
manifestation of the CPT degeneracy described in
Eqs.~\eqref{eq:osc-deg} and~\eqref{eq:NSI-deg} once $\Dmq_{21}$ and
$\Eps_\oplus'$ are set to zero~\cite{GonzalezGarcia:2011my}.

As further simplification, in order to keep the fit manageable we
assume real NSI, which we implement by choosing $\alpha_1 = \alpha_2 =
0$ with $\varphi_{ij}$ range $-\pi/2 \leq \varphi_{ij} \leq \pi/2$.
It is important to note that with these approximations the formalism
for atmospheric and LBL data is CP-conserving. We will go back to this
point when discussing the experimental results included in the
analysis.

In addition to atmospheric and LBL experiments, important information
on neutrino oscillation parameters is provided also by reactor
experiments with a baseline of about 1~km. Due to the very small
amount of matter crossed, both standard and non-standard matter
effects are completely irrelevant for these experiments, and the
corresponding $P_{ee}$ survival probability depends only on the vacuum
parameters. However, in view of the high precision recently attained
by both reactor and LBL experiments in the determination of the
atmospheric mass-squared difference, combining them without adopting a
full $3\nu$ oscillation scheme requires a special care. In
Ref.~\cite{Nunokawa:2005nx} it was shown that, in the limit $\Dmq_{21}
\ll \Dmq_{31}$ as indicated by the data, the $P_{\mu\mu}$ probability
relevant for LBL-disappearance experiments can be accurately described
in terms of a single effective mass parameter $\Dmq_{\mu\mu} =
\Dmq_{31} - r_2 \Dmq_{21}$ with $r_2 = |U_{\mu2}^\text{vac}|^2 \big/
(|U_{\mu1}^\text{vac}|^2 + |U_{\mu2}^\text{vac}|^2)$. In the rest of
this work we will therefore make use of $\Dmq_{\mu\mu}$ as the
fundamental quantity parametrizing the atmospheric mass-squared
difference. For each choice of the vacuum mixing parameters in
$U_\text{vac}$, the calculations for the various data sets are then
performed as follows:
\begin{itemize}
\item for atmospheric and LBL data we assume $\Dmq_{21}=0$ and set
  $\Dmq_{31} = \Dmq_{\mu\mu}$;
  
\item for reactor neutrinos we keep $\Dmq_{21}$ finite and set
  $\Dmq_{31} = \Dmq_{\mu\mu} + r_2 \Dmq_{21}$.
\end{itemize}
In this way the information provided by reactor and long-baseline data
on the atmospheric mass scale is consistently combined in spite of the
approximation $\Dmq_{21} \to 0$ discussed above. Note that the
correlations between solar and reactor neutrinos are properly taken
into account in our fit, in particular for what concerns the octant of
$\theta_{12}$.

\subsection{Matter potential for solar and KamLAND neutrinos}

For the study of propagation of solar and KamLAND neutrinos one can
work in the one mass dominance approximation, $\Dmq_{31} \to \infty$
(which effectively means that $G_F \sum_f N_f(x) \Eps_{\alpha\beta}^f
\ll \Dmq_{31} / E_\nu$). In this approximation the survival
probability $P_{ee}$ can be written as~\cite{Kuo:1986sk, Guzzo:2000kx}
\begin{equation}
  \label{eq:peesun}
  P_{ee} = c_{13}^4 P_\text{eff} + s_{13}^4
\end{equation}
The probability $P_\text{eff}$ can be calculated in an effective
$2\times 2$ model described by the Hamiltonian $H_\text{eff} =
H_\text{vac}^\text{eff} + H_\text{mat}^\text{eff}$, with:
\begin{align}
  \label{eq:HvacSol}
  H_\text{vac}^\text{eff}
  &= \frac{\Dmq_{21}}{4 E_\nu}
  \begin{pmatrix}
    -\cos2\theta_{12} \, \hphantom{e^{-i\delta_\text{CP}}}
    & ~\sin2\theta_{12} \, e^{i\delta_\text{CP}}
    \\
    \hphantom{-}\sin2\theta_{12} \, e^{-i\delta_\text{CP}}
    & ~\cos2\theta_{12} \, \hphantom{e^{i\delta_\text{CP}}}
  \end{pmatrix} ,
  \\
  \label{eq:HmatSol}
  H_\text{mat}^\text{eff}
  &= \sqrt{2} G_F N_e(x)
  \left[
    \begin{pmatrix}
      c_{13}^2 & 0 \\
      0 & 0
    \end{pmatrix}
    + \big[ \xi^p + Y_n(x) \xi^n \big]
    \begin{pmatrix}
      -\Eps_D^{\eta\hphantom{*}} & \Eps_N^\eta \\
      \hphantom{+} \Eps_N^{\eta*} & \Eps_D^\eta
    \end{pmatrix}
    \right],
\end{align}
where we have imposed the quark-lepton factorization of
Eq.~\eqref{eq:epx-eta} and used the parametrization convention of
Eq.~\eqref{eq:Uvac} for $U_\text{vac}$.  The coefficients
$\Eps_D^\eta$ and $\Eps_N^\eta$ are related to the original parameters
$\Eps_{\alpha\beta}^\eta$ by the following relations:
\begin{align}
  \label{eq:eps_D}
  \begin{split}
    \Eps_D^\eta
    &= c_{13} s_{13}\, \Re\!\big( s_{23} \, \Eps_{e\mu}^\eta
    + c_{23} \, \Eps_{e\tau}^\eta \big)
    - \big( 1 + s_{13}^2 \big)\, c_{23} s_{23}\,
    \Re\!\big( \Eps_{\mu\tau}^\eta \big)
    \\
    & \hphantom{={}}
    -\frac{c_{13}^2}{2} \big( \Eps_{ee}^\eta - \Eps_{\mu\mu}^\eta \big)
    + \frac{s_{23}^2 - s_{13}^2 c_{23}^2}{2}
    \big( \Eps_{\tau\tau}^\eta - \Eps_{\mu\mu}^\eta \big) \,,
  \end{split}
  \\[2mm]
  \label{eq:eps_N}
  \Eps_N^\eta &=
  c_{13} \big( c_{23} \, \Eps_{e\mu}^\eta - s_{23} \, \Eps_{e\tau}^\eta \big)
  + s_{13} \left[
    s_{23}^2 \, \Eps_{\mu\tau}^\eta - c_{23}^2 \, \Eps_{\mu\tau}^{\eta*}
    + c_{23} s_{23} \big( \Eps_{\tau\tau}^\eta - \Eps_{\mu\mu}^\eta \big)
    \right].
\end{align}
Note that the $\delta_\text{CP}$ phase appearing in
Eq.~\eqref{eq:HvacSol} could be transferred to Eq.~\eqref{eq:HmatSol}
without observable consequences by means of a global rephasing. Hence,
for each fixed value of $\eta$ the relevant probabilities for solar
and KamLAND neutrinos depend effectively on six quantities: the three
real oscillation parameters $\Dmq_{21}$, $\theta_{12}$ and
$\theta_{13}$, one real matter parameter $\Eps_D^\eta$, and one
complex vacuum-matter combination $\Eps_N^\eta
e^{-i\delta_\text{CP}}$.
As stated in Sec.~\ref{sec:formalism-earth} in this work we will
assume real NSI, implemented here by setting $\delta_\text{CP} = 0$
and considering only real (both positive and negative) values for
$\Eps_N^\eta$.

Unlike in the Earth, the matter chemical composition of the Sun varies
substantially along the neutrino trajectory, and consequently the
potential depends non-trivially on the specific combinations of
couplings with up and down quarks --- \textit{i.e.}, on the value of
$\eta$. This implies that the generalized mass-ordering degeneracy is
not exact, except for $\eta=0$ (in which case the NSI potential is
proportional to the standard MSW potential and an exact inversion of
the matter sign is possible). However, as we will see in
Sec.~\ref{sec:solar}, the CPT transformation described in
Eqs.~\eqref{eq:osc-deg} and~\eqref{eq:NSI-deg} still results in a good
fit to the global analysis of oscillation data for a wide range of
values of $\eta$, and non-oscillation data are needed to break this
degeneracy~\cite{Coloma:2017egw, Coloma:2017ncl}. Because of the
change in the $\theta_{12}$ octant implied by Eq.~\eqref{eq:osc-deg}
and given that the standard LMA solution clearly favors $\theta_{12} <
45^\circ$, this alternative solution is characterized by a value of
$\theta_{12} > 45^\circ$. In what follows we will denote it as
``LMA-D''~\cite{Miranda:2004nb}.

\section{Analysis of solar and KamLAND data}
\label{sec:solar}

Let us start by presenting the results of the updated analysis of
solar and KamLAND experiments in the context of oscillations with the
generalized matter potential in Eq.~\eqref{eq:HmatSol}.  For KamLAND
we include the separate DS1, DS2, DS3 spectra~\cite{Gando:2013nba}
with reactor fluxes as determined by Daya-Bay~\cite{An:2016srz}.  In
the analysis of solar neutrino data we consider the total rates from
the radiochemical experiments Chlorine~\cite{Cleveland:1998nv},
Gallex/GNO~\cite{Kaether:2010ag} and SAGE~\cite{Abdurashitov:2009tn},
the results for the four phases of
Super-Kamiokande~\cite{Hosaka:2005um, Cravens:2008aa, Abe:2010hy,
  sksol:nakano2016} (including the 2055 days separate day and night
spectra from Ref.~\cite{sksol:nakano2016} of Super-Kamiokande IV), the
combined data of the three phases of SNO as presented in
Ref.~\cite{Aharmim:2011vm}, and the results of both Phase-I and
Phase-II of Borexino~\cite{Bellini:2011rx, Bellini:2008mr,
  Bellini:2014uqa}.

\begin{pagefigure}\centering
  \includegraphics[width=0.93\textwidth]{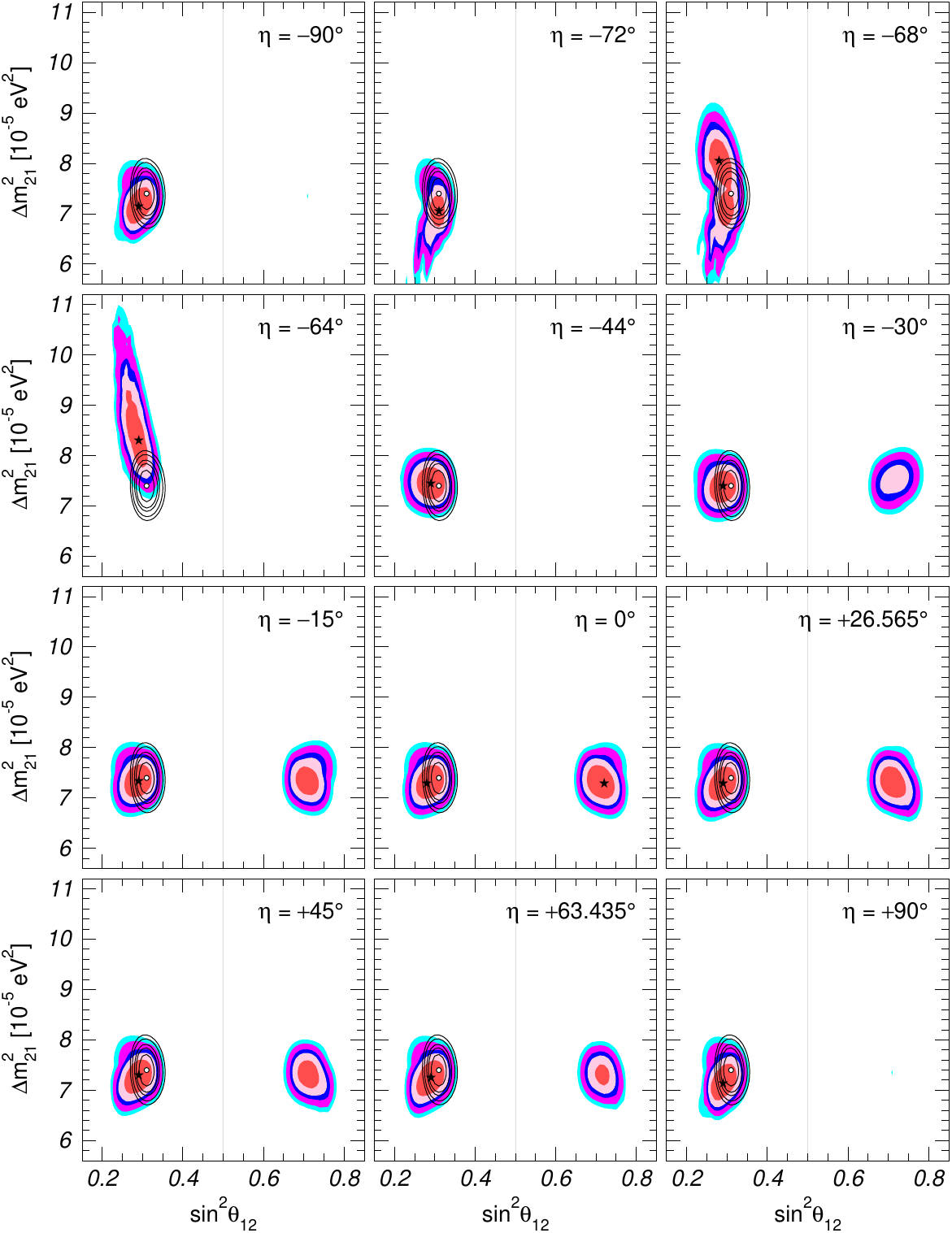}
  \caption{Two-dimensional projections of the $1\sigma$, 90\%,
    $2\sigma$, 99\% and $3\sigma$ CL (2~dof) allowed regions from the
    analysis of solar and KamLAND data in the presence of non-standard
    matter potential for the oscillation parameters $(\theta_{12},
    \Dmq_{21})$ after marginalizing over the NSI parameters and for
    $\theta_{13}$ fixed to $\sin^2\theta_{13} = 0.022$.  The best-fit
    point is marked with a star. The results are shown for fixed
    values of the NSI quark coupling parameter $\eta$. For comparison
    the corresponding allowed regions for the analysis in terms of
    $3\nu$ oscillations without NSI are shown as black void
    contours. Note that, as a consequence of the periodicity of
    $\eta$, the regions in the first ($\eta = -90^\circ$) and last
    ($\eta = +90^\circ$) panels are identical.}
  \label{fig:sun-oscil}
\end{pagefigure}

\begin{figure}\centering
  \includegraphics[width=0.8\textwidth]{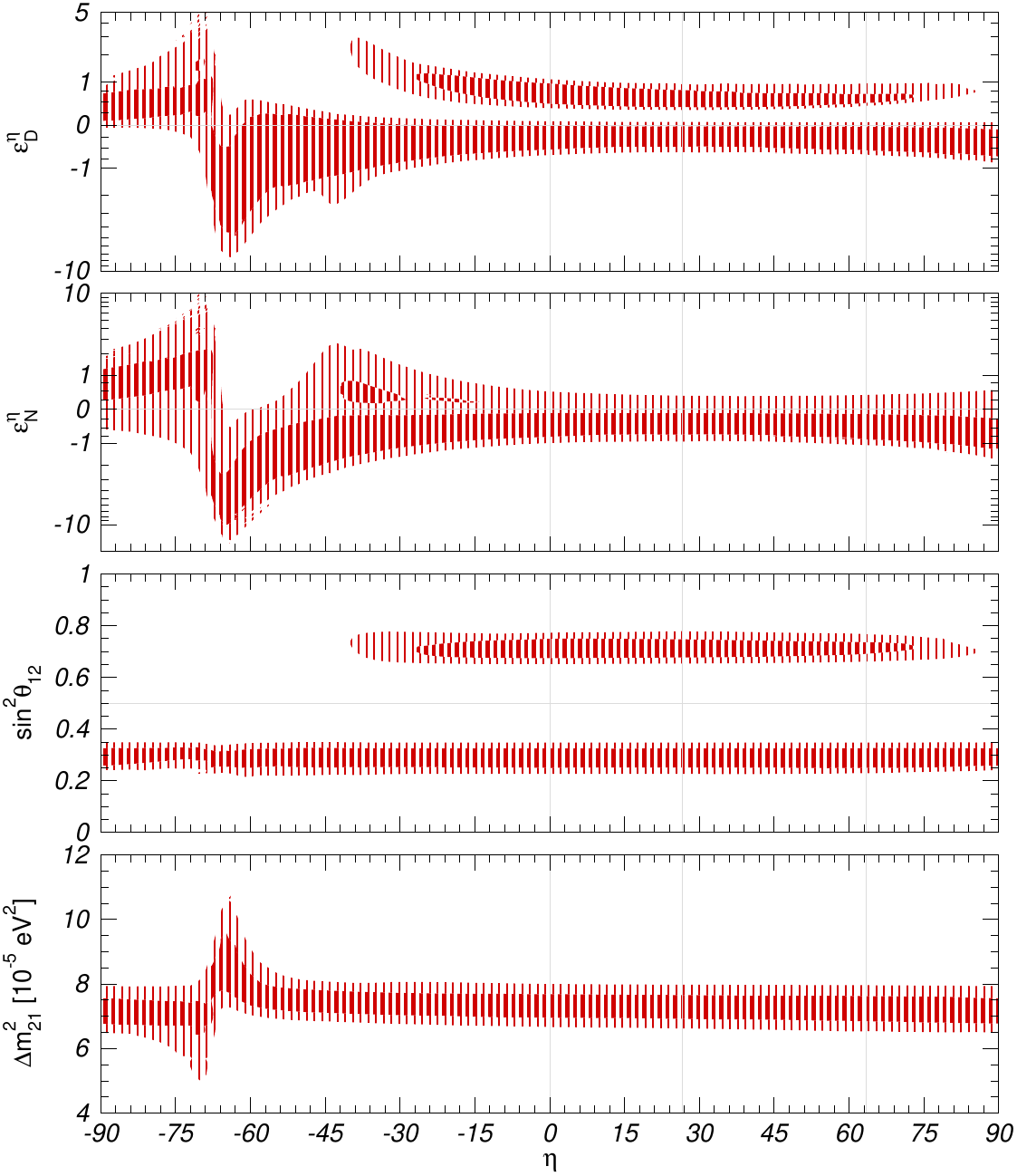}
  \caption{90\% and $3\sigma$ CL (1~dof) allowed ranges from the
    analysis of solar and KamLAND data in the presence of non-standard
    neutrino-matter interactions, for the four relevant parameters
    (the matter potential parameters $\Eps_D^\eta$ and $\Eps_N^\eta$
    as well as the oscillation parameters $\Dmq_{21}$ and
    $\sin^2\theta_{12}$) as a function of the NSI quark coupling
    parameter $\eta$, for $\sin^2\theta_{13}=0.022$.  In each panel
    the three undisplayed parameters have been marginalized.}
    \label{fig:sun-range}
\end{figure}

We present different projections of the allowed parameter space in
Figs.~\ref{fig:sun-oscil}--\ref{fig:sun-range}. In the analysis we
have fixed $\sin^2\theta_{13} = 0.022$ which is the best-fit value
from the global analysis of $3\nu$ oscillations~\cite{Esteban:2016qun,
  nufit-3.2}.\footnote{Note that the determination of $\theta_{13}$ is
  presently dominated by reactor experiments, which have negligible
  matter effects and are therefore unaffected by the presence of NSI.
  Allowing for variations of $\theta_{13}$ within its current
  well-determined range has no quantitative impact on our results.} So
for each value of $\eta$ there are four relevant parameters:
$\Dmq_{21}$, $\sin^2\theta_{12}$, $\Eps_D^\eta$, and $\Eps_N^\eta$.
As mentioned above, for simplicity the results are shown for real
$\Eps_N^\eta$. Also strictly speaking the sign of $\Eps_N^\eta$ is not
physically observable in oscillation experiments, as it can be
reabsorbed into a redefinition of the sign of $\theta_{12}$. However,
for definiteness we have chosen to present our results in the
convention $\theta_{12} \geq 0$, and therefore we consider both
positive and negative values of $\Eps_N^\eta$.
Fig.~\ref{fig:sun-oscil} shows the two-dimensional projections on the
oscillation parameters $(\theta_{12}, \Dmq_{21})$ for different values
of $\eta$ after marginalizing over the NSI parameters, while
Fig.~\ref{fig:sun-epses} shows the corresponding two-dimensional
projections on the matter potential parameters $(\Eps_D^\eta,
\Eps_N^\eta)$ after marginalizing over the oscillation
parameters. The one-dimensional ranges for the four parameters as a
function of $\eta$ are shown in Fig.~\ref{fig:sun-range}.

\begin{pagefigure}\centering
  \includegraphics[width=0.85\textwidth]{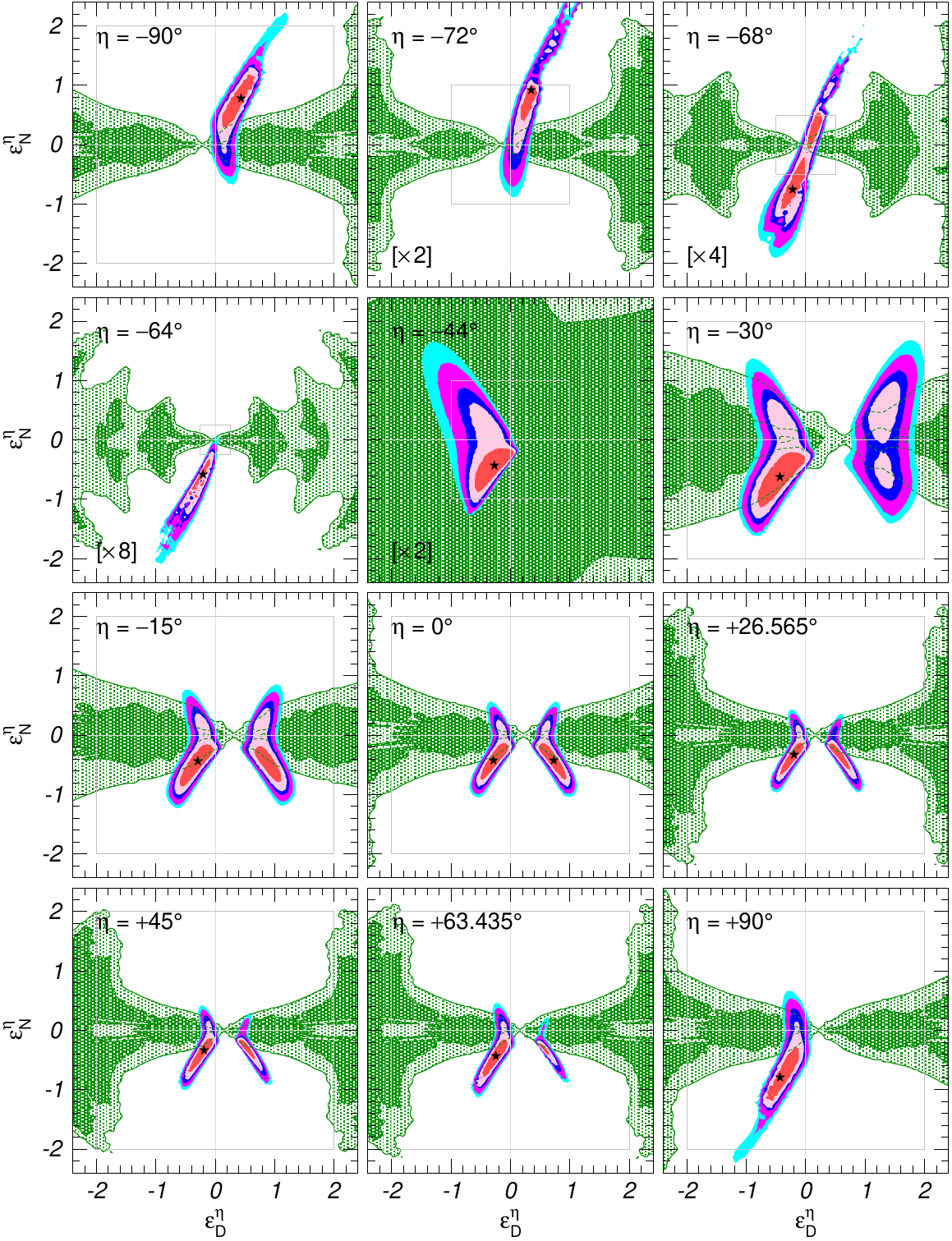}
  \caption{Two-dimensional projections of the $1\sigma$, 90\%,
    $2\sigma$, 99\% and $3\sigma$ CL (2~dof) allowed regions from the
    analysis of solar and KamLAND data in the presence of non-standard
    matter potential for the matter potential parameters
    $(\Eps_D^\eta, \Eps_N^\eta)$, for $\sin^2\theta_{13} = 0.022$ and
    after marginalizing over the oscillation parameters.  The best-fit
    point is marked with a star. The results are shown for fixed
    values of the NSI quark coupling parameter $\eta$.  The panels
    with a scale factor ``$[\times N]$'' in their lower-left corner
    have been ``zoomed-out'' by such factor with respect to the
    standard axis ranges, hence the grey square drawn in each panel
    always corresponds to $\max\big( |\Eps_D^\eta|, |\Eps_N^\eta|
    \big) = 2$ and has the same size in all the panels.  For
    illustration we also show as shaded green areas the 90\% and
    $3\sigma$ CL allowed regions from the analysis of the atmospheric
    and LBL data. Note that, as a consequence of the periodicity of
    $\eta$, the regions in the first ($\eta = -90^\circ$) and last
    ($\eta = +90^\circ$) panels are identical up to an overall sign
    flip.}
  \label{fig:sun-epses}
\end{pagefigure}

The first thing to notice in the figures is the presence of the LMA-D
solution for a wide range of values of $\eta$. This is a consequence
of the approximate degeneracy discussed in the previous section. In
particular, as expected, for $\eta=0$ the degeneracy is exact and the
LMA-D region in Fig.~\ref{fig:sun-oscil} is perfectly symmetric to the
LMA one with respect to maximal $\theta_{12}$.  Looking at the
corresponding panels of Fig.~\ref{fig:sun-epses} we note that the
allowed area in the NSI parameter space is composed by two
disconnected regions, one containing the SM case (\textit{i.e.}, the
point $\Eps_D^\eta = \Eps_N^\eta = 0$) which corresponds to the
``standard'' LMA solution in the presence of the modified matter
potential, and another which does not include such point and
corresponds to the LMA-D solution.  Although the appearance of the
LMA-D region is a common feature, there is also a range of values of
$\eta$ for which such solution is strongly disfavored and does not
appear at the displayed CL's.

\begin{figure}\centering
  \includegraphics[width=0.9\textwidth]{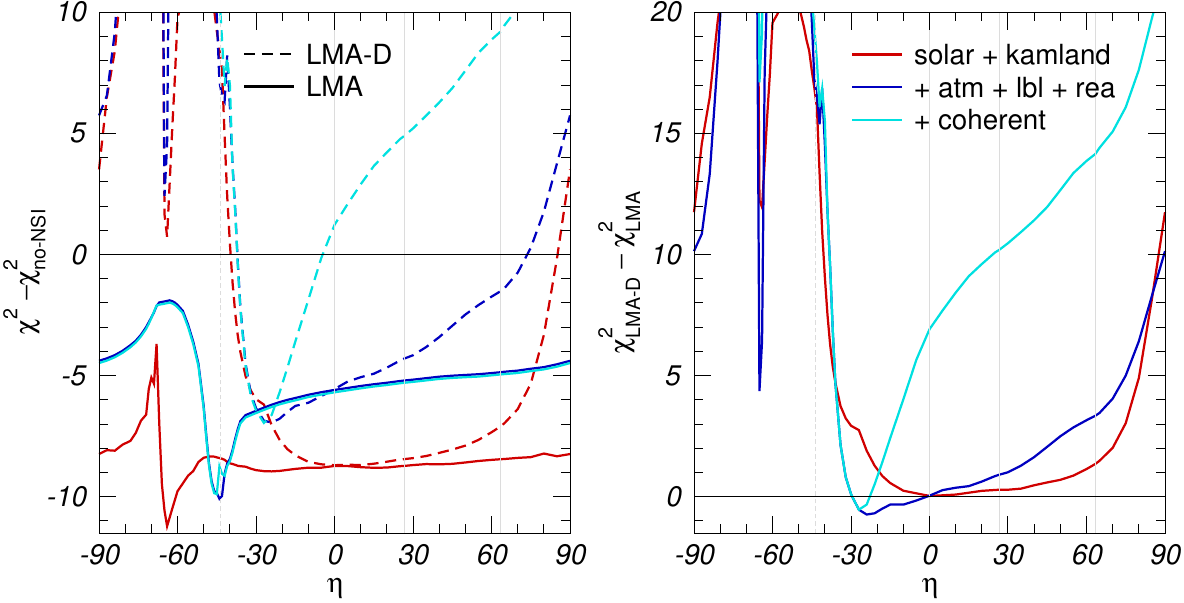}
  \caption{Left: $\chi^2_\text{LMA}(\eta) - \chi^2_\text{no-NSI}$
    (full lines) and $\chi^2_\text{LMA-D}(\eta) -
    \chi^2_\text{no-NSI}$ (dashed lines) for the analysis of different
    data combinations (as labeled in the figure) as a function of the
    NSI quark coupling parameter $\eta$.  Right:
    $\chi^2_\text{LMA-D}(\eta) - \chi^2_\text{LMA}(\eta)$ as a
    function of $\eta$. See text for details.}
  \label{fig:chisq-eta}
\end{figure}

In order to further illustrate the $\eta$ dependence of the results,
it is convenient to introduce the functions $\chi^2_\text{LMA}(\eta)$
and $\chi^2_\text{LMA-D}(\eta)$ which are obtained by marginalizing
the $\chi^2$ for a given value of $\eta$ over both the oscillation and
the matter potential parameters with the constraint $\theta_{12} <
45^\circ$ and $\theta_{12} > 45^\circ$, respectively. With this, in
the left panel of Fig.~\ref{fig:chisq-eta} we plot the differences
$\chi^2_\text{LMA}(\eta) - \chi^2_\text{no-NSI}$ (full lines) and
$\chi^2_\text{LMA-D}(\eta) - \chi^2_\text{no-NSI}$ (dashed lines),
where $\chi^2_\text{no-NSI}$ is the minimum $\chi^2$ for standard
$3\nu$ oscillations (\textit{i.e.}, without NSI), while in the right
panel we plot $\chi^2_\text{LMA-D}(\eta) - \chi^2_\text{LMA}(\eta)$
which quantifies the relative quality of the LMA and LMA-D solutions.
From this plot we can see that even for the analysis of solar and
KamLAND data alone (red lines) the LMA-D solution is disfavored at
more than $3\sigma$ when $\eta \lesssim -40^\circ$ or $\eta \gtrsim
86^\circ$. Generically for such range of $\eta$ the modified matter
potential in the Sun, which in the presence of NSI is determined not
only by the density profile but also by the chemical composition, does
not allow for a degenerate solution compatible with KamLAND data. In
particular, as discussed below, for a fraction of those $\eta$ values
the NSI contribution to the matter potential in the Sun becomes very
suppressed and therefore the degeneracy between NSI and octant of
$\theta_{12}$ cannot be realized. In what respects the LMA solution,
we notice that it always provides a better fit (or equivalent for
$\eta=0$) than the LMA-D solution to solar and KamLAND data, for any
value of $\eta$. This does not have to be the case in general, and
indeed it is no longer so when atmospheric data are also included in
the analysis. We will go back to this point in the next section.

From the left panel in Fig.~\ref{fig:chisq-eta} we see that the
introduction of NSI can lead to a substantial improvement in the
analysis of solar and KamLAND data, resulting in a sizable decrease
of the minimum $\chi^2$ with respect to the standard oscillation
scenario.  The maximum gain occur for $\eta \simeq -64^\circ$ and is
about $11.2$ units in $\chi^2$ (\textit{i.e.}, a $3.3\sigma$ effect),
although for most of the values of $\eta$ the inclusion of NSI
improves the combined fit to solar and KamLAND by about $2.5\sigma$.
This is mainly driven by the well known tension between solar and
KamLAND data in the determination of $\Dmq_{21}$. The phenomenological
status of such tension has not changed significantly over the last
lustrum, and arises essentially from a combination of two effects: (a)
the $^8$B measurements performed by SNO, SK and Borexino does not show
any evidence of the low energy spectrum turn-up expected in the
standard LMA-MSW~\cite{Wolfenstein:1977ue, Mikheev:1986gs} solution
for the value of $\Dmq_{21}$ favored by KamLAND, and (b) the
observation of a non-vanishing day-night asymmetry in SK, whose size
is considerably larger than what predicted for the $\Dmq_{21}$ value
indicated of KamLAND. With the data included in the analysis this
results into a tension of $\Delta\chi^2\sim 7.4$ for the standard
$3\nu$ oscillations.
Such tension can be alleviated in presence of a non-standard matter
potential, thus leading to the corresponding decrease in the minimum
$\chi^2$ for most values of $\eta$ --- with the exception of the range
$-70^\circ \lesssim \eta \lesssim -60^\circ$.  Furthermore, as seen in
the lower panel in Fig.~\ref{fig:sun-range} the allowed range of
$\Dmq_{21}$ implied by the combined solar and KamLAND data is pretty
much independent of the specific value of $\eta$, except again for
$-70^\circ \lesssim \eta\lesssim -60^\circ$ in which case it can
extend well beyond the standard oscillation LMA values.

The special behaviour of the likelihood of solar and KamLAND in the
range $-70^\circ \lesssim \eta\lesssim -60^\circ$ is a consequence of
the fact that for such values the NSI contributions to the matter
potential in the Sun approximately cancel. As mentioned in the
previous section, the matter chemical composition of the Sun varies
substantially along the neutrino production region, with $Y_n(x)$
dropping from about $1/2$ in the center to about $1/6$ at the border
of the solar core. Thus for $-70^\circ \lesssim \eta \lesssim
-60^\circ$ (corresponding to $-2.75 \lesssim \tan\eta \lesssim -1.75$)
the effective NSI couplings $\Epx_{\alpha\beta}(x) =
\Eps_{\alpha\beta}^p + Y_n(x) \Eps_{\alpha\beta}^n \propto 1 + Y_n(x)
\tan\eta \to 0$ vanish at some point inside the neutrino production
region.  This means that for such values of $\eta$ the constraints on
the NSI couplings from solar data become very weak, being prevented
from disappearing completely only by the \emph{gradient} of
$Y_n(x)$. This is visible in the two upper panels in
Fig.~\ref{fig:sun-range} and in the panels of Fig.~\ref{fig:sun-epses}
with $\eta$ in such range, where a multiplicative factor 2--8 has to
be included to make the regions fit in the same axis range.  Indeed
for those values of $\eta$ the allowed NSI couplings can be so large
that their effect in the propagation of long-baseline reactor
neutrinos through the Earth becomes sizable, and can therefore lead
to spectral distortions in KamLAND which affect the determination of
$\Dmq_{21}$ --- hence the ``migration'' and distortion of the LMA
region observed in the corresponding panels in
Fig.~\ref{fig:sun-oscil}.  In particular, it is precisely for $\eta =
-64^\circ$ for which the ``migration'' of the KamLAND region leads to
the best agreement with the solar determination of $\Dmq_{12}$,
whereas for $\eta = -68^\circ$ we find the worst agreement.  In any
case, looking at the shaded green regions in the corresponding panels
of Fig.~\ref{fig:sun-epses} we can anticipate that the inclusion of
atmospheric and LBL oscillation experiments will rule out almost
completely such very large NSI values.

As for $\theta_{12}$, looking at the relevant panel in
Fig.~\ref{fig:sun-range} we can see that its determination is pretty
much independent of the value of $\eta$, however a comparison between
colored and void regions in Fig.~\ref{fig:sun-oscil} shows that its
allowed range always extends to lower values than in the standard
$3\nu$ case without NSI. This is expected since the presence of
non-diagonal NSI parametrized by $\Eps_N^\eta$ provides another source
of flavor transition, thus leading to a weakening of the lower bound
on $\theta_{12}$.

We finish this section by noticing that two of the panels in
Figs.~\ref{fig:sun-oscil} and~\ref{fig:sun-epses} correspond to the
values of NSI only with $f=u$ ($\eta \approx 26.6^\circ$) and only
with $f=d$ ($\eta \approx 63.4^\circ$) and can be directly compared
with the results of our previous global OSC+NSI analysis in
Ref.~\cite{Gonzalez-Garcia:2013usa}.  For illustration we also show in
one of the panels the results for $\eta = -44^\circ$ which is close to
the value for which NSI effects in the Earth matter cancel.

\section{Results of the global oscillation analysis}
\label{sec:globalosc}

In addition to the solar and KamLAND data discussed so far, in our
global analysis we also consider the following data sets:
\begin{itemize}
\item atmospheric neutrino data: this sample includes the four phases
  of Super-Kamiokande (up to 1775 days of SK4~\cite{Wendell:2014dka})
  in the form of the ``classical'' samples of $e$-like and $\mu$-like
  events (70 energy and zenith angle bins), together with the complete
  set of DeepCore 3-year $\mu$-like events (64 data points) presented
  in Ref.~\cite{Aartsen:2014yll} and publicly released in
  Ref.~\cite{deepcore:2016}. The calculations of the event rates for
  both detectors are based on the atmospheric neutrino flux
  calculations described in Ref.~\cite{Honda:2015fha}. In addition, we
  also include the results on $\nu_\mu$-induced upgoing muons reported
  by IceCube~\cite{Jones:2015, Arguelles:2015, TheIceCube:2016oqi},
  based on one year of data taking;
  
\item long-baseline experiments: we include here the $\nu_\mu$ and
  $\bar\nu_\mu$ disappearance as well as the $\nu_e$ and $\bar\nu_e$
  appearance data in MINOS~\cite{Adamson:2013whj} (39, 14, 5, and 5
  data points, respectively), the $\nu_\mu$ and $\bar\nu_\mu$
  disappearance data in T2K~\cite{t2k:vietnam2016} (39 and 55 data
  points, respectively), and the $\nu_\mu$ disappearance data in
  NO$\nu$A~\cite{nova:fnal2018} (72 data points).  As mentioned in
  Sec.~\ref{sec:formalism}, in order to keep the fit manageable we
  restrict ourselves to the CP-conserving scenario. At present, the
  results of the full $3\nu$ oscillation analysis with standard matter
  potential show a hint of CP violation~\cite{Esteban:2016qun,
    nufit-3.2}, which is mainly driven by the LBL $\nu_e$ and
  $\bar\nu_e$ appearance data at T2K~\cite{t2k:vietnam2016} and
  NO$\nu$A~\cite{nova:fnal2018}.  Conversely, allowing for CP
  violation has negligible impact on the determination of the
  CP-conserving parameters in the analysis of MINOS appearance data
  and of any LBL disappearance data samples, as well as in our
  analysis of atmospheric events mentioned above.  Hence, to ensure
  full consistency with our CP-conserving parametrization we have
  chosen \emph{not} to include in the present study the data from the
  $\nu_e$ and $\bar\nu_e$ appearance channels in NO$\nu$A and
  T2K. This also renders our fit only marginally sensitive to the
  neutrino mass ordering. In what follows we will refer to the
  long-baseline data included here as LBL-CPC.  Note that for
  simplicity we have omitted from our analysis the MINOS+ results on
  $\nu_\mu$ disappearance, despite the fact that they probe higher
  neutrino energies than the other LBL experiments and are therefore,
  at least in principle, more sensitive to the NSI parameters than,
  \textit{e.g.}, MINOS~\cite{Graf:2015egk}.  The rationale behind this
  choice is that the LBL experiments which we include are crucial to
  determine the oscillation parameters in an energy range where NSI
  effects are subdominant, whereas at present MINOS+ data lack this
  capability. As for the NSI parameters involved in $\nu_\mu$
  disappearance, they are more strongly constrained by the atmospheric
  neutrino data of SK and IceCube, which extends to energies well
  beyond those of MINOS+;
  
\item medium-baseline (MBL) reactor experiments: since these
  experiments are largely insensitive to matter effects (either
  standard or non-standard), the results included here coincide with
  those of the standard $3\nu$ analysis presented in
  Ref.~\cite{nufit-3.2} and illustrated in the black lines of the plot
  tagged <<Synergies:~determination of $\Dmq_{3\ell}$>>. Such analysis
  is based on a reactor-flux-independent approach as described in
  Ref.~\cite{Dentler:2017tkw}, and includes the Double-Chooz FD-I/ND
  and FD-II/ND spectral ratios with 455-day (FD-I), 363-day (FD-II),
  and 258-day (ND) exposures~\cite{dc:cabrera2016} (56 data points),
  the Daya-Bay 1230-day EH2/EH1 and EH3/EH1 spectral
  ratios~\cite{An:2016ses} (70 data points), and the Reno 1500-day
  FD/ND spectral ratios~\cite{reno:eps2017} (26 data points).
\end{itemize}

\begin{figure}\centering
  \includegraphics[width=\textwidth]{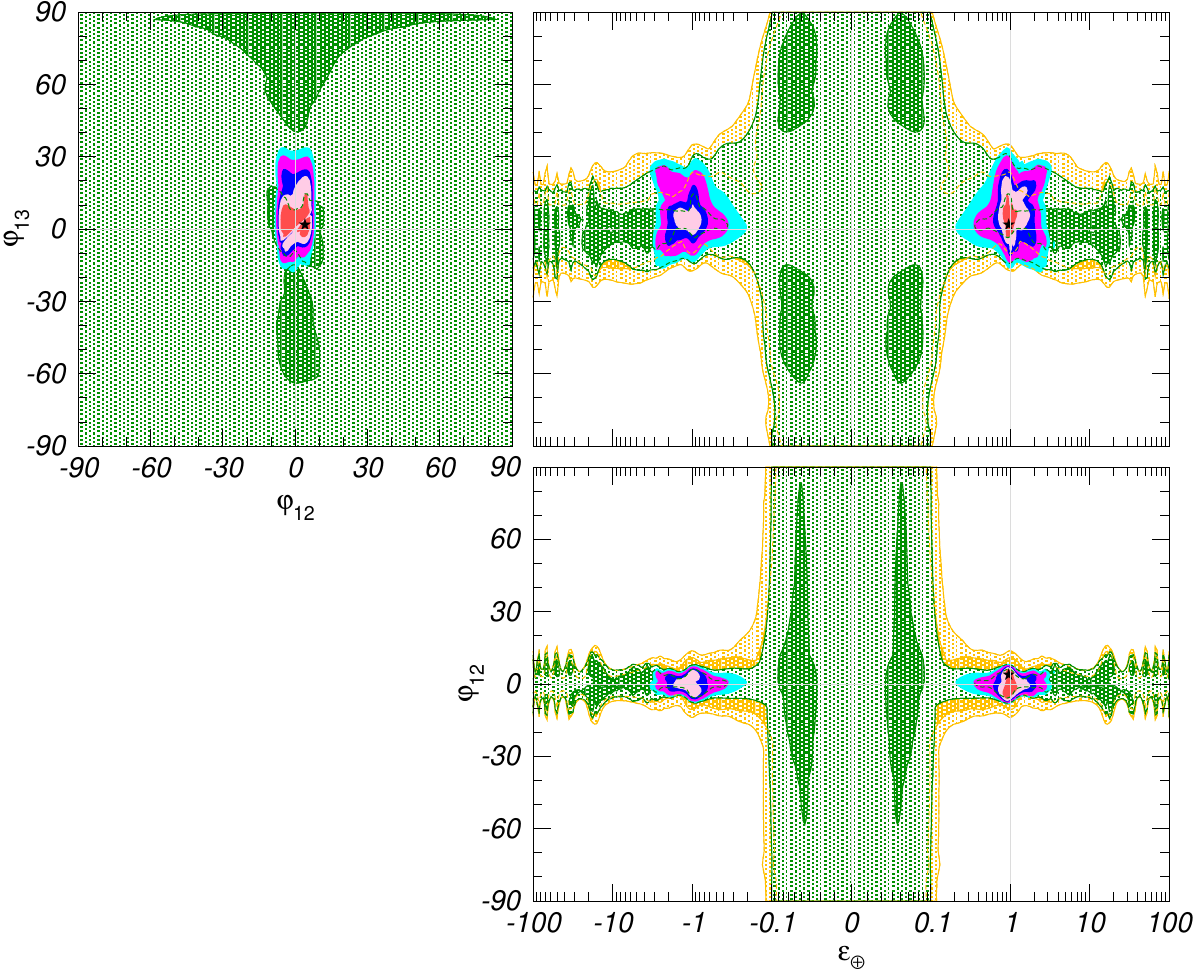}
  \caption{Two-dimensional projections of the allowed regions onto the
    matter potential parameters $\Eps_\oplus$, $\varphi_{12}$, and
    $\varphi_{13}$ after marginalization with respect to the
    undisplayed parameters. The large green regions correspond to the
    analysis of atmospheric, LBL-CPC, and MBL reactor data at 90\% and
    $3\sigma$ CL. For comparison we show in yellow the corresponding
    results when omitting IceCube and reactor data. The solid colored
    regions show the $1\sigma$, 90\%, $2\sigma$, 99\% and $3\sigma$ CL
    allowed regions once solar and KamLAND data are included. The
    best-fit point is marked with a star.}
  \label{fig:glb-epsil}
\end{figure}

Let us begin by showing in Figure~\ref{fig:glb-epsil} the
two-dimensional projections of the allowed regions in the Earth's
matter potential parameters $\Eps_\oplus$, $\varphi_{12}$ and
$\varphi_{13}$ (\textit{i.e.}, in the parametrization of
Eq.~\eqref{eq:eps_atm} with $\alpha_i = 0$) after marginalizing over
the oscillation parameters.  The green regions show the 90\% and
$3\sigma$ confidence regions (2~dof) from the analysis of atmospheric,
LBL-CPC and MBL reactor experiments.  Besides the increase in
statistics on low-energy atmospheric events provided by the updated
Super-Kamiokande and the new DeepCore data samples, the main
difference with respect to the analysis in
Refs.~\cite{GonzalezGarcia:2011my, Gonzalez-Garcia:2013usa} is the
inclusion of the bounds on NSI-induced $\nu_\mu$ disappearance
provided by IceCube high-energy data as well as the precise
information on $\theta_{13}$ and $|\Dmq_{31}|$ from MBL reactor
experiments.  To illustrate their impact we show as yellow regions the
results obtained when IceCube and reactor data are omitted. For what
concerns the projection over the matter potential parameters shown
here, we have verified that the difference between the yellow and
green regions is mostly driven by IceCube, which restricts the allowed
values of the $\varphi_{12}$ for $|\Eps_\oplus| \sim 0.1$--$1$.  This
can be understood since, for neutrino with energies above
$\mathcal{O}(100~\text{GeV})$, the vacuum oscillation is very
suppressed and the survival probability of atmospheric $\nu_\mu$
arriving at zenith angle $\Theta_\nu$ is dominated by the matter
induced transitions
\begin{equation}
  P_{\mu\mu} \simeq 1 - \sin^2( 2\varphi_{\mu\mu} )
  \sin^2\left( \frac{d_e(\Theta_\nu) \Eps_\oplus}{2} \right)
  \quad\text{with}\quad
  \sin^2\varphi_{\mu\mu} = \sin^2\varphi_{12} \cos^2\varphi_{13}
\end{equation}
where $d_e(\Theta_\nu) = \sqrt{2} G_F X_e(\Theta_\nu)$ and the column
density $X_e(\Theta_\nu)$ is the integral of $N_e(x)$ along the
neutrino path in the Earth~\cite{Gonzalez-Garcia:2016gpq}. Since
$0.2\lesssim d_e(\Theta_\nu) \lesssim 20$ for $-1 \leq \cos\Theta_\nu
\leq -0.2$, the range $0.1 \lesssim |\Eps_\oplus| \lesssim 1$
corresponds to the first oscillation maximum for some of the
trajectories.  Also, the effective parameter $\varphi_{\mu\mu}$
entering in the expression of $P_{\mu\mu}$ depends linearly on
$\varphi_{12}$ and only quadratically on $\varphi_{13}$, which
explains why the bounds on the mixings are stronger for $\varphi_{12}$
than for $\varphi_{13}$.

As can be seen in Fig.~\ref{fig:glb-epsil}, even with the inclusion of
IceCube neither upper nor lower bounds on the overall strength of the
Earth's matter effects, $\Eps_\oplus$, can be derived from the
analysis of atmospheric, LBL-CPC and MBL reactor
experiments~\cite{Friedland:2004ah, Friedland:2005vy,
  GonzalezGarcia:2011my}.\footnote{See Refs.~\cite{Esmaili:2013fva,
    Salvado:2016uqu} for constraints in more restricted NSI
  scenarios.}  This happens because the considered data sample is
mainly sensitive to NSI through $\nu_\mu$ disappearance, and lacks
robust constraints on matter effects in the $\nu_e$ sector.  As a
consequence, when marginalizing over $\Eps_\oplus$ (as well as over
the oscillation parameters) the full flavor projection $(\varphi_{12},
\varphi_{13})$ plane is allowed.  On the other hand, once the results
of solar and KamLAND experiments (which are sensitive to $\nu_e$) are
included in the analysis a bound on $\Eps_\oplus$ is obtained and the
flavor structure of the matter potential in the Earth is significantly
constrained.

\begin{figure}\centering
  \includegraphics[width=\textwidth]{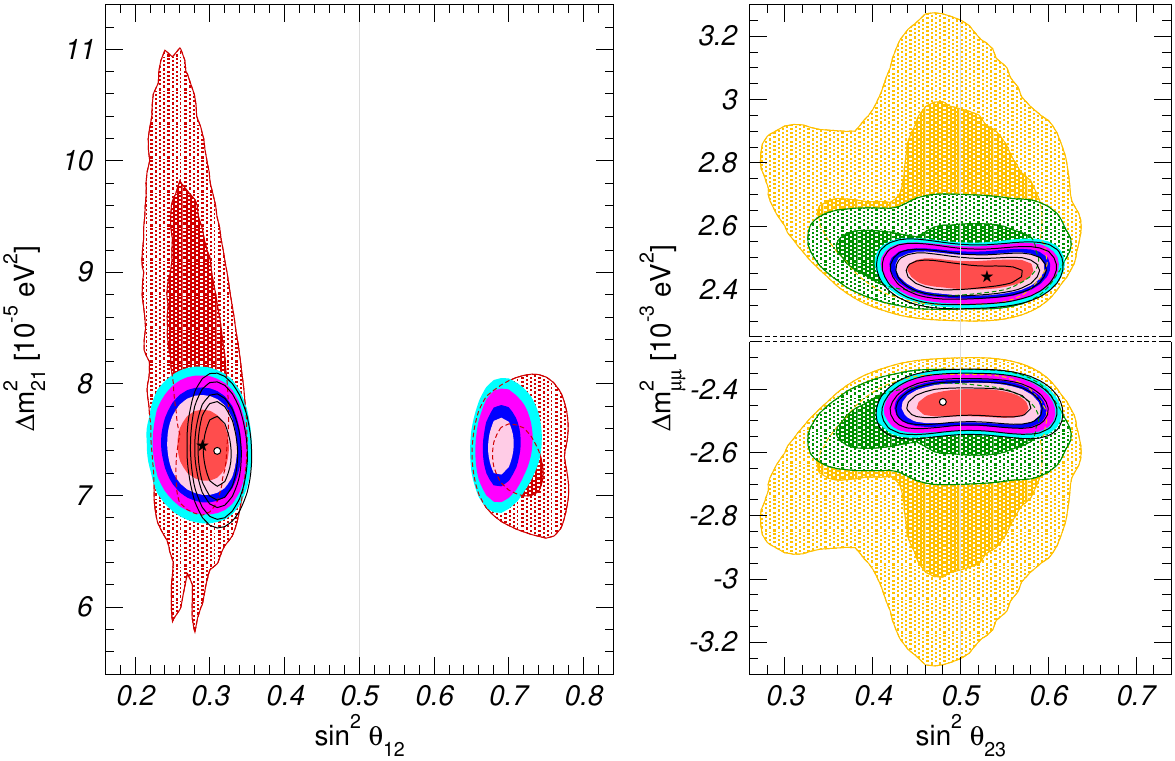}
  \caption{Two-dimensional projections of the allowed regions onto
    different vacuum parameters after marginalizing over the matter
    potential parameters (including $\eta$) and the undisplayed
    oscillation parameters.  The solid colored regions correspond to
    the global analysis of all oscillation data, and show the
    $1\sigma$, 90\%, $2\sigma$, 99\% and $3\sigma$ CL allowed regions;
    the best-fit point is marked with a star.  The black void regions
    correspond to the analysis with the standard matter potential
    (\textit{i.e.}, without NSI) and its best-fit point is marked with
    an empty dot.  For comparison, in the left panel we show in red
    the 90\% and $3\sigma$ allowed regions including only solar and
    KamLAND results, while in the right panels we show in green the
    90\% and $3\sigma$ allowed regions excluding solar and KamLAND
    data, and in yellow the corresponding ones excluding also IceCube
    and reactor data.}
  \label{fig:glb-oscil}
\end{figure}

In Fig.~\ref{fig:glb-oscil} we show the two-dimensional projections of
the allowed regions from the global analysis onto different sets of
oscillation parameters.  These regions are obtained after
marginalizing over the undisplayed vacuum parameters as well as the
NSI couplings.  For comparison we also show as black-contour void
regions the corresponding results with the standard matter potential,
\textit{i.e.}, in the absence of NSI.  As discussed in
Sec.~\ref{sec:formalism-earth}, in the right panels we have chosen to
plot the regions in terms of the effective mass-squared difference
relevant for $\nu_\mu$ disappearance experiments,
$\Dmq_{\mu\mu}$. Notice that, having omitted NO$\nu$A and T2K
appearance data and also set $\Dmq_{21} = 0$ in atmospheric and
LBL-CPC experiments, the impact of the mass ordering on the results of
the fit is greatly reduced.

This figure clearly shows the robustness of the determination of the
$\Dmq_{21}$, $|\Dmq_{\mu\mu}|$ and $\theta_{23}$ vacuum oscillation
parameters even in the presence of the generalized NSI
interactions. This result relies on the complementarity and synergies
between the different data sets, which allows to constrain those
regions of the parameter space where cancellations between standard
and non-standard effects occur in a particular data set.  To
illustrate this we show as shaded regions the results obtained when
some of the data are removed.  For example, comparing the solid
colored regions with the shaded red ones in the left panel we see how,
in the presence of NSI with arbitrary values of $\eta$, the precise
determination of $\Dmq_{21}$ requires the inclusion of atmospheric,
LBL-CPC and MBL reactor data: if these sets are omitted, the huge
values of the NSI couplings allowed by solar data for $-70^\circ
\lesssim \eta \lesssim -60^\circ$ destabilize KamLAND's determination
of $\Dmq_{21}$, as discussed in Sec.~\ref{sec:solar}.  The inclusion
of these sets also limits the margins for NSI to alleviate the tension
between solar and KamLAND data on the preferred $\Dmq_{21}$ value, as
can be seen by comparing the full dark-blue and red lines in the left
panel of Fig.~\ref{fig:chisq-eta}: indeed, in the global analysis the
best-fit is achieved for $\eta \simeq -44^\circ$, which is precisely
when the NSI effects in the Earth matter cancel so that no restriction
on NSI contributions to solar and KamLAND data is imposed.

In the same way we see on the right panels that, if the solar and
KamLAND data are removed from the fit, the determination of
$\Dmq_{\mu\mu}$ and $\theta_{23}$ degrades because of the possible
cancellations between NSI and mass oscillation effects in the relevant
atmospheric and LBL-CPC probabilities. As NSI lead to
energy-independent contributions to the oscillation phase, such
cancellations allow for larger values of $|\Dmq_{\mu\mu}|$.  Comparing
the yellow and green regions we see the inclusion MBL reactor
experiments, for which NSI effects are irrelevant due to the short
baselines involved, is crucial to reduce the degeneracies and provide
a NSI-independent measurement of $|\Dmq_{\mu\mu}|$. Even so, only the
inclusion of solar and KamLAND allows to recover the full sensitivity
of atmospheric and LBL-CPC experiments and derive limits on
$\Dmq_{\mu\mu}$ and $\theta_{23}$ as robust as the standard ones.

The most dramatic implications of NSI for what concerns the
determination of the oscillation parameters affect $\theta_{12}$.  In
particular, for generic NSI with arbitrary $\eta$ the LMA-D solution
is still perfectly allowed by the global oscillation analysis, as
indicated by the presence of the corresponding region in the left
panel in Fig.~\ref{fig:glb-oscil}.  Turning to
Fig.~\ref{fig:chisq-eta} we see that even after including all the
oscillation data (dark-blue lines) the LMA-D solution is allowed at
$3\sigma$ for $-38^\circ \lesssim \eta \lesssim 87^\circ$ (as well as
in a narrow window around $\eta \simeq -65^\circ$), and indeed for
$-28^\circ \lesssim \eta\lesssim 0^\circ$ it provides a slightly
better global fit than LMA.
From Fig.~\ref{fig:glb-oscil} we also see that the lower bound on
$\theta_{12}$ in the presence of NSI is substantially weaker than the
standard $3\nu$ case.  We had already noticed such reduction in the
analysis of solar and KamLAND data for any value of $\eta$; here we
point out that the cancellation of matter effects in the Earth for
$\eta \approx -43.6^\circ$ prevents any improvement of that limit from
the addition of Earth-based oscillation experiments.

\begin{figure}\centering
  \includegraphics[width=0.8\textwidth]{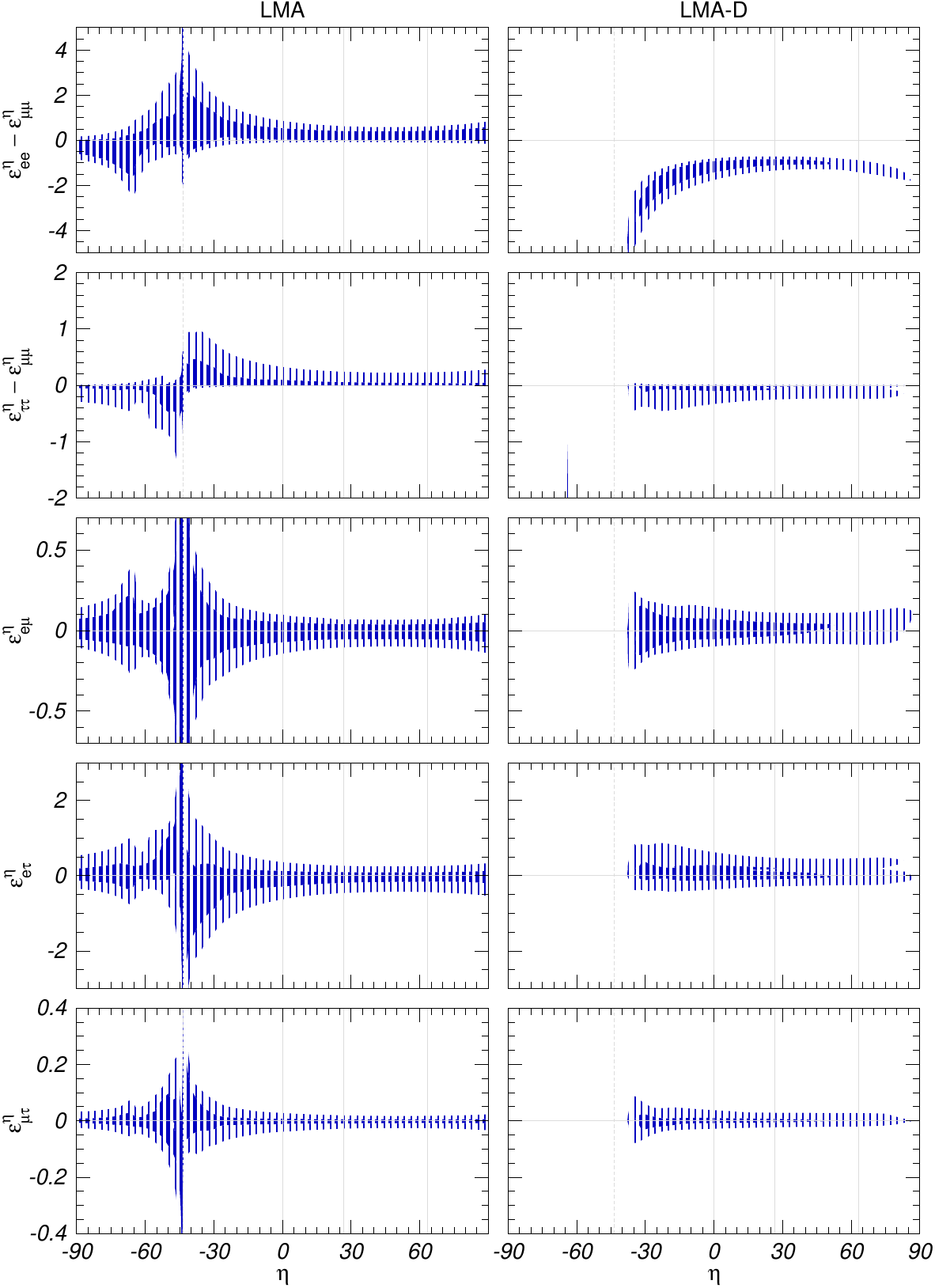}
  \caption{90\%, and $3\sigma$ CL (1~dof) allowed ranges for the NSI
    couplings from the global oscillation analysis in the presence of
    non-standard matter potential as a function of the NSI quark
    coupling parameter $\eta$.  In each panel the undisplayed
    parameters have been marginalized. On the left panels the
    oscillation parameters have been marginalized within the LMA
    region while the right panels corresponds to LMA-D solutions.  The
    ranges are defined with respect to the minimum for each $\eta$.}
  \label{fig:glb-range}
\end{figure}

The bounds on the five relevant NSI couplings (two diagonal
differences and three non-diagonal entries) from the global
oscillation analysis are displayed in Fig.~\ref{fig:glb-range} as a
function of $\eta$. Concretely, for each value of $\eta$ we plot as
vertical bars the 90\% and $3\sigma$ allowed ranges (1~dof) after
marginalizing with respect to the undisplayed parameters. The left and
right panels correspond to the limits for $\theta_{12}$ within the LMA
and LMA-D solution, respectively, both defined with respect to the
same common minimum for each given $\eta$.  For the sake of
convenience and comparison with previous results we list in the first
columns in Table~\ref{tab:ranges} the 95\% CL ranges for NSI with
up-quarks only ($\eta \approx 26.6^\circ$), down-quarks only ($\eta
\approx 63.4^\circ$) and couplings proportional to the electric charge
($\eta=0^\circ$); in this last case we have introduced an extra
$\sqrt{5}$ normalization factor so that the quoted bounds can be
directly interpreted in terms of $\Eps_{\alpha\beta}^p$.  Let us point
out that the sign of each non-diagonal $\Eps_{\alpha\beta}^\eta$ can
be flipped away by a suitable change of signs in some of the mixing
angles; it is therefore not an intrinsic property of NSI, but rather a
relative feature of the vacuum and matter Hamiltonians.  Thus,
strictly speaking, once the results are marginalized with respect to
all the other parameters in the most general parameter space, the
oscillation analysis can only provide bounds on
$|\Eps_{\alpha\neq\beta}^\eta|$.  However, for definiteness we have
chosen to restrict the range of the mixing angles to $0 \leq
\theta_{ij} \leq \pi/2$ and to ascribe the relative vacuum-matter
signs to the NSI couplings, so that the ranges of the non-diagonal
$\Eps_{\alpha\beta}^\eta$ in Figs.~\ref{fig:glb-range}
and~\ref{fig:coh-range} as well as in Table~\ref{tab:ranges} are given
for both signs.

From Fig.~\ref{fig:glb-range} and Table~\ref{tab:ranges} we see that
the allowed range for all the couplings (except $\Eps_{ee}^\eta -
\Eps_{\mu\mu}^\eta$) obtained marginalizing over both $\theta_{12}$
octants, which we denote in the table as $\text{LMA} \oplus
\text{LMA-D}$, is only slighter wider than what obtained considering
only the LMA solution. Conversely, for $\Eps_{ee}^\eta -
\Eps_{\mu\mu}^\eta$ the allowed range is composed by two disjoint
intervals, each one corresponding to a different $\theta_{12}$
octant. Note that for this coupling the interval associated with the
LMA solution is not centered at zero due to the tension between the
value of $\Dmq_{21}$ preferred by KamLAND and solar experiments, even
after including the bounds from atmospheric and long-baseline data.
In general, we find that the allowed ranges for all the couplings do
not depend strongly on the value of $\eta$ as long as $\eta$ differs
enough from the critical value $\eta \approx -43.6^\circ$. As already
explained, at this point non-standard interactions in the Earth cancel
out, so that no bound on the NSI parameters can be derived from any
Earth-based experiment. This leads to a breakdown of the limits on
$\Eps_{\alpha\beta}^\eta$, since solar data are only sensitive to the
$\Eps_D^\eta$ and $\Eps_N^\eta$ combinations and cannot constrain the
five NSI couplings simultaneously.  In addition to the region around
$\eta \approx -43.6^\circ$, there is also some mild weakening of the
bounds on NSI couplings involving $\nu_e$ for $-70^\circ \lesssim \eta
\lesssim -60^\circ$, corresponding to the window where NSI effects in
the Sun are suppressed.  Apart from these special cases, the bounds
quoted in Table~\ref{tab:ranges} are representative of the
characteristic sensitivity to the NSI coefficients from present
oscillation experiments, which at 95\% CL ranges from
$\mathcal{O}(1\%)$ for $|\Eps_{\mu\tau}^\eta|$ to $\mathcal{O}(30\%)$
for $|\Eps_{e\tau}^\eta|$ ---~the exception being, of course,
$\Eps_{ee}^\eta - \Eps_{\mu\mu}^\eta$.

\begin{table}\centering
  \definecolor{grey}{gray}{0.75}
  \newcommand{\grsep}{~\color{grey}\vrule}
  \begin{tabular}{|l@{\grsep}r@{\grsep}r|l@{\grsep}r@{\grsep}r|}
    \hline
    \multicolumn{3}{|c|}{OSC}
    & \multicolumn{3}{c|}{+COHERENT}
    \\
    \hline
    & {LMA\hfil} & $\text{LMA}\oplus\text{LMA-D}$ &
    & {LMA\hfil} & $\text{LMA}\oplus\text{LMA-D}$
    \\
    \hline
    \begin{tabular}{@{}l@{}}
      $\Eps_{ee}^u - \Eps_{\mu\mu}^u$ \\
      $\Eps_{\tau\tau}^u - \Eps_{\mu\mu}^u$
    \end{tabular}
    &
    \begin{tabular}{@{}r@{}}
      $[-0.020, +0.456]$ \\
      $[-0.005, +0.130]$
    \end{tabular}
    &
    \begin{tabular}{@{}r@{}}
      $\oplus [-1.192, -0.802]$ \\
      $[-0.152, +0.130]$
    \end{tabular}
    &
    \begin{tabular}{@{}l@{}}
      $\Eps_{ee}^u$ \\
      $\Eps_{\mu\mu}^u$ \\
      $\Eps_{\tau\tau}^u$
    \end{tabular}
    &
    \begin{tabular}{@{}r@{}}
      $[-0.008, +0.618]$ \\
      $[-0.111, +0.402]$ \\
      $[-0.110, +0.404]$
    \end{tabular}
    &
    \begin{tabular}{@{}r@{}}
      $[-0.008, +0.618]$ \\
      $[-0.111, +0.402]$ \\
      $[-0.110, +0.404]$
    \end{tabular}
    \\
    $\Eps_{e\mu}^u$ & $[-0.060, +0.049]$ & $[-0.060, +0.067]$ &
    $\Eps_{e\mu}^u$ & $[-0.060, +0.049]$ & $[-0.060, +0.049]$
    \\
    $\Eps_{e\tau}^u$ & $[-0.292, +0.119]$ & $[-0.292, +0.336]$ &
    $\Eps_{e\tau}^u$ & $[-0.248, +0.116]$ & $[-0.248, +0.116]$
    \\
    $\Eps_{\mu\tau}^u$ & $[-0.013, +0.010]$ & $[-0.013, +0.014]$ &
    $\Eps_{\mu\tau}^u$ & $[-0.012, +0.009]$ & $[-0.012, +0.009]$
    \\
    \hline
    \begin{tabular}{@{}l@{}}
      $\Eps_{ee}^d - \Eps_{\mu\mu}^d$ \\
      $\Eps_{\tau\tau}^d - \Eps_{\mu\mu}^d$
    \end{tabular}
    &
    \begin{tabular}{@{}r@{}}
      $[-0.027, +0.474]$ \\
      $[-0.005, +0.095]$
    \end{tabular}
    &
    \begin{tabular}{@{}r@{}}
      $\oplus [-1.232, -1.111]$ \\
      $[-0.013, +0.095]$
    \end{tabular}
    &
    \begin{tabular}{@{}l@{}}
      $\Eps_{ee}^d$ \\
      $\Eps_{\mu\mu}^d$ \\
      $\Eps_{\tau\tau}^d$
    \end{tabular}
    &
    \begin{tabular}{@{}r@{}}
      $[-0.012, +0.565]$ \\
      $[-0.103, +0.361]$ \\
      $[-0.102, +0.361]$
    \end{tabular}
    &
    \begin{tabular}{@{}r@{}}
      $[-0.012, +0.565]$ \\
      $[-0.103, +0.361]$ \\
      $[-0.102, +0.361]$
    \end{tabular}
    \\
    $\Eps_{e\mu}^d$ & $[-0.061, +0.049]$ & $[-0.061, +0.073]$ &
    $\Eps_{e\mu}^d$ & $[-0.058, +0.049]$ & $[-0.058, +0.049]$
    \\
    $\Eps_{e\tau}^d$ & $[-0.247, +0.119]$ & $[-0.247, +0.119]$ &
    $\Eps_{e\tau}^d$ & $[-0.206, +0.110]$ & $[-0.206, +0.110]$
    \\
    $\Eps_{\mu\tau}^d$ & $[-0.012, +0.009]$ & $[-0.012, +0.009]$ &
    $\Eps_{\mu\tau}^d$ & $[-0.011, +0.009]$ & $[-0.011, +0.009]$
    \\
    \hline
    \begin{tabular}{@{}l@{}}
      $\Eps_{ee}^p - \Eps_{\mu\mu}^p$ \\
      $\Eps_{\tau\tau}^p - \Eps_{\mu\mu}^p$
    \end{tabular}
    &
    \begin{tabular}{@{}r@{}}
      $[-0.041, +1.312]$ \\
      $[-0.015, +0.426]$
    \end{tabular}
    &
    \begin{tabular}{@{}r@{}}
      $\oplus [-3.327, -1.958]$ \\
      $[-0.424, +0.426]$
    \end{tabular}
    &
    \begin{tabular}{@{}l@{}}
      $\Eps_{ee}^p$ \\
      $\Eps_{\mu\mu}^p$ \\
      $\Eps_{\tau\tau}^p$
    \end{tabular}
    &
    \begin{tabular}{@{}r@{}}
      $[-0.010, +2.039]$ \\
      $[-0.364, +1.387]$ \\
      $[-0.350, +1.400]$
    \end{tabular}
    &
    \begin{tabular}{@{}r@{}}
      $[-0.010, +2.039]$ \\
      $[-0.364, +1.387]$ \\
      $[-0.350, +1.400]$
    \end{tabular}
    \\
    $\Eps_{e\mu}^p$ & $[-0.178, +0.147]$ & $[-0.178, +0.178]$ &
    $\Eps_{e\mu}^p$ & $[-0.179, +0.146]$ & $[-0.179, +0.146]$
    \\
    $\Eps_{e\tau}^p$ & $[-0.954, +0.356]$ & $[-0.954, +0.949]$ &
    $\Eps_{e\tau}^p$ & $[-0.860, +0.350]$ & $[-0.860, +0.350]$
    \\
    $\Eps_{\mu\tau}^p$ & $[-0.035, +0.027]$ & $[-0.035, +0.035]$ &
    $\Eps_{\mu\tau}^p$ & $[-0.035, +0.028]$ & $[-0.035, +0.028]$
    \\
    \hline
  \end{tabular}
  \caption{$2\sigma$ allowed ranges for the NSI couplings
    $\Eps_{\alpha\beta}^u$, $\Eps_{\alpha\beta}^d$ and
    $\Eps_{\alpha\beta}^p$ as obtained from the global analysis of
    oscillation data (left column) and also including COHERENT
    constraints.  The results are obtained after marginalizing over
    oscillation and the other matter potential parameters either
    within the LMA only and within both LMA and LMA-D subspaces
    respectively (this second case is denoted as $\text{LMA} \oplus
    \text{LMA-D}$). Notice that once COHERENT data are included the
    two columns become identical, since for NSI couplings with
    $f=u,d,p$ the LMA-D solution is only allowed well above 95\% CL.}
  \label{tab:ranges}
\end{table}

\section{Combined analysis of oscillation and COHERENT data}
\label{sec:coherent}

To conclude our study, let us now quantify the impact of adding to our
fit the constraints on coherent neutrino--nucleus scattering from the
first results of the COHERENT experiment~\cite{Akimov:2017ade}.  As
discussed in the introduction, while the bounds from oscillation
effects apply to models where the NSI are generated by mediators of
arbitrarily light masses, for scattering experiments there is a
minimum mediator mass below which the contact interaction
approximation is not adequate to describe the $\nu$ interactions in
the detector. This threshold can be estimated by noticing that if the
NC-NSI are generated by the exchange of a mediator of mass $M$ with
characteristic coupling to fermions $g$, then $\Eps_{\alpha\beta}^f
G_F \sim \mathcal{O}(g^2/M^2)$ which can give a correction to the
number of coherent scattering events (for NSI couplings interfering
with the SM) of the order $N_\text{NSI} / N_\text{SM} \sim
g^2/(q^2+M^2) \, (1/G_F) \sim \Eps_{\alpha\beta}^f M^2/(Q^2+M^2)$,
with $Q^2$ being the characteristic momentum transfer in the
scattering. For COHERENT we have $Q^2\sim (50~\text{MeV})^2$ so that
$\Eps_{\alpha\beta}^f \sim 0.1-1$ can lead to a 5\% effect if $M
\gtrsim \mathcal{O}(10-50~\text{MeV}).$\footnote{This naive estimate
  agrees well with the range of mediators obtained in the detailed
  analysis of COHERENT bounds in a $Z'$ model performed in
  Ref.~\cite{Denton:2018xmq}.}  Hence the bounds presented here apply
for models for which the mediator responsible for the NSI is heavier
than about 10~MeV.

\begin{figure}\centering
  \includegraphics[width=0.8\textwidth]{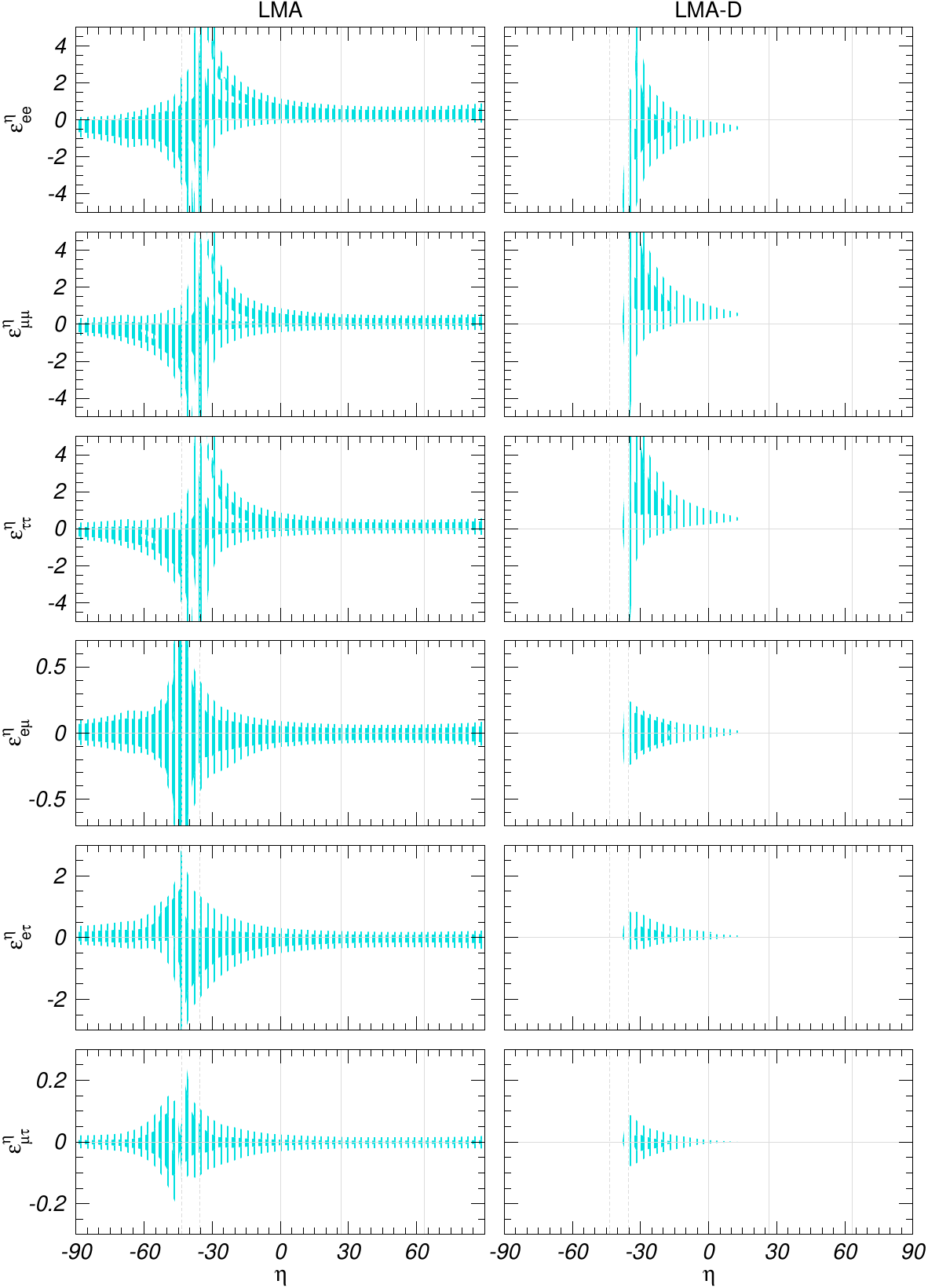}
  \caption{90\% and $3\sigma$ CL (1~dof) allowed ranges for the NSI
    couplings from the global oscillation analysis in the presence of
    non-standard matter potential combines with COHERENT as a function
    of the NSI quark coupling parameter $\eta$. In each panel the
    undisplayed parameters have been marginalized. On the left panels
    the oscillation parameters have been marginalized within the LMA
    region while the right panels corresponds to LMA-D solutions.  The
    ranges are defined with respect to the minimum for each $\eta$.}
  \label{fig:coh-range}
\end{figure}

For the statistical analysis of the COHERENT results we follow
Ref.~\cite{Coloma:2017ncl} and construct $\chi^2_\text{COH}$ using
just the total number of events, according to the expression given in
the supplementary material of Ref.~\cite{Akimov:2017ade}.  The
predicted number of signal events $N_\text{NSI}$ can be expressed as:
\begin{equation}
  N_\text{NSI} = \gamma
  \left[ f_{\nu_e} Q_{w e}^2 + (f_{\nu_\mu} + f_{\bar\nu_\mu})Q_{w \mu}^2 \right] \,,
\end{equation}
where $\gamma$ is an overall normalization constant, the coefficients
$f_{\nu_e} = 0.31$, $f_{\nu_\mu} = 0.19$, and $f_{\bar\nu_\mu} = 0.50$
are the relative contributions from the three flux components
($\nu_e$, $\nu_\mu$ and $\bar\nu_\mu$), and the terms $Q_{w\alpha}^2$
encode the dependence on the NSI couplings:
\begin{equation}
  \label{eq:Qw}
  Q_{w\alpha}^2 \propto \sum_i
  \bigg\lbrace
  \big[ Z_i (g_p^V + \Eps_{\alpha\alpha}^p)
    + N_i (g_n^V +\Eps_{\alpha\alpha}^n) \big]^2
  + \sum_{\beta\neq\alpha}
  \big[ Z_i \Eps_{\alpha\beta}^p + N_i \Eps_{\alpha\beta}^n \big]^2
  \bigg\rbrace
\end{equation}
where we have used the effective matrices $\Eps_{\alpha\beta}^p$ and
$\Eps_{\alpha\beta}^n$ defined in Eq.~\eqref{eq:eps-nucleon}. In this
expression $i \in \lbrace \text{Cs}, \text{I} \rbrace$ is the sum over
the target nuclei, $Z_i$ and $N_i$ are the corresponding number of
protons and neutrons ($Z_\text{Cs} = 55$, $N_\text{Cs} = 78$ for
cesium and $Z_\text{I} = 53$, $N_\text{I} = 74$ for iodine), and
$g_p^V = 1/2 - 2\sin^2\theta_W$ and $g_n^V = -1/2$ are the SM vector
couplings of the $Z$ boson to protons and neutrons, respectively, with
$\theta_W$ being the weak mixing angle. Note that the neutron/proton
ratio in the two target nuclei is very similar, $N_\text{Cs} \big/
Z_\text{Cs} \simeq 1.419$ for cesium and $N_\text{I} \big/ Z_\text{I}
\simeq 1.396$ for iodine, with an average value $Y_n^\text{coh} =
1.407$. We can therefore approximate Eq.~\eqref{eq:Qw} as:
\begin{equation}
  Q_{w\alpha}^2 \propto \big[
    (g_p^V + Y_n^\text{coh} g_n^V) + \Eps_{\alpha\alpha}^\text{coh} \big]^2
  + \sum_{\beta\neq\alpha}
  \big( \Eps_{\alpha\beta}^\text{coh} \big)^2
  \quad\text{with}\quad
  \Eps_{\alpha\beta}^\text{coh}
  \equiv \Eps_{\alpha\beta}^p + Y_n^\text{coh} \Eps_{\alpha\beta}^n \,.
\end{equation}
After imposing quark-lepton factorization from Eq.~\eqref{eq:epx-eta},
$\Eps_{\alpha\beta}^\text{coh}$ can be written as:
\begin{equation}
  \Eps_{\alpha\beta}^\text{coh}
  = \Eps_{\alpha\beta}^\eta \big( \xi^p +Y_n^\text{coh} \xi^n \big)
  = \sqrt{5} \left( \cos\eta + Y_n^\text{coh} \sin\eta \right)
  \Eps_{\alpha\beta}^\eta \,.
\end{equation}
This expression is formally identical to Eq.~\eqref{eq:eps-earth},
except for the numerical value of $Y_n$. This suggests that the
analysis of Earth-based oscillation experiments and of coherent
scattering data share a number of phenomenological features. In
particular, the best-fit value and allowed ranges of
$\Eps_{\alpha\beta}^\text{coh}$ implied by COHERENT are independent of
$\eta$, while the corresponding bounds on the physical quantities
$\Eps_{\alpha\beta}^\eta$ simply scale as $(\cos\eta + Y_n^\text{coh}
\sin\eta)$. Also, for $\eta = \arctan(-1/Y_n^\text{coh}) \approx
-35.4^\circ$ no bound on $\Eps_{\alpha\beta}^\eta$ can be derived from
COHERENT data.

The results of the global analysis of oscillation plus COHERENT data
are shown as cyan lines in Fig.~\ref{fig:chisq-eta}; the corresponding
ranges for the NSI coefficients are shown in Fig.~\ref{fig:coh-range}
and in the right column of Table~\ref{tab:ranges}.  As can be seen,
the main impact of including COHERENT data is to strongly disfavor the
LMA-D solution for a wide range of $\eta$.  LMA-D is allowed below
$3\sigma$ only for $-38^\circ \lesssim \eta \lesssim 14^\circ$.  This
generalizes the results of Ref.~\cite{Coloma:2017ncl} to a wider set
of NSI-NC with quarks.  We also find that the allowed ranges of flavor
non-diagonal NSI couplings are moderately reduced.  More
interestingly, the addition of COHERENT data allows to derive
constraints on each of the diagonal parameters separately.  This is
especially relevant for $\Eps_{\tau\tau}^\eta$ for which the bounds
become more than an order of magnitude stronger than previous indirect
(loop induced) limits~\cite{Davidson:2003ha} for most $\eta$ values.
We notice, however, that COHERENT data are still not strong enough to
disfavor the large ranges of NSI allowed by oscillations for $\eta
\approx -43.6^\circ$. Moreover, the cancellation of NSI effects in
COHERENT data for $\eta \approx -35.4^\circ$ implies that no separate
reconstruction of the diagonal parameters is possible around such
value.

\begin{figure}\centering
  \includegraphics[width=\textwidth]{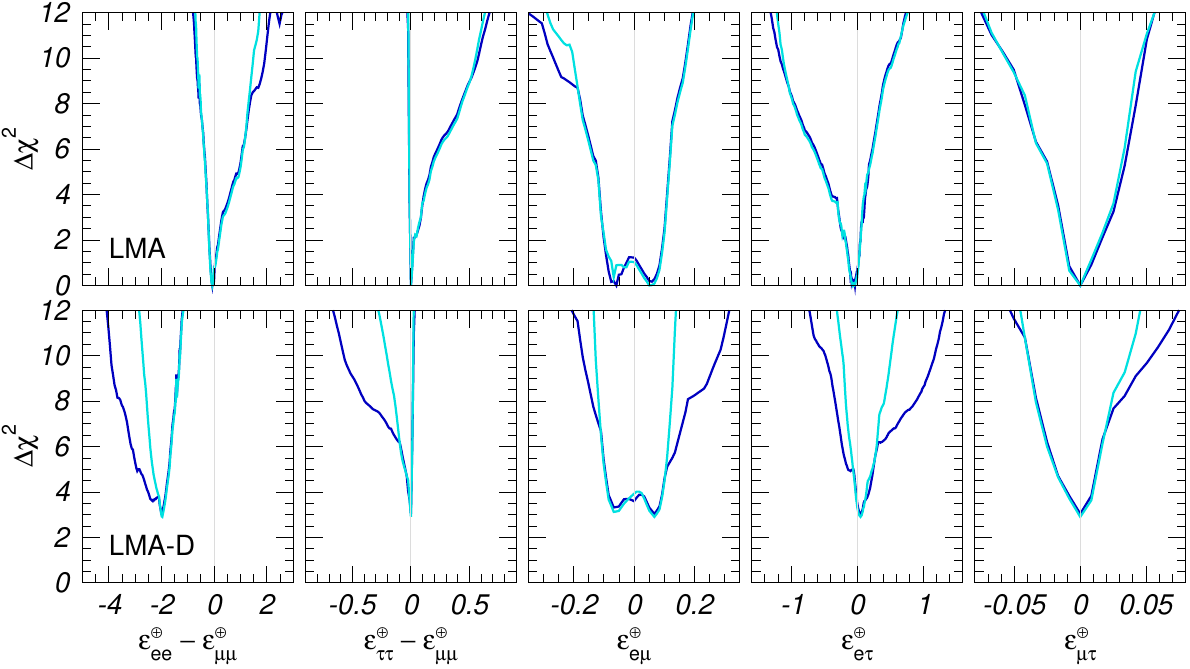}
  \caption{Dependence of the $\Delta\chi^2$ function on the effective
    NSI parameters relevant for matter effects in LBL experiments with
    arbitrary values of $\eta$, from the global analysis of solar,
    atmospheric, LBL-CPC and reactor data (blue lines) and including
    also COHERENT (cyan lines).  The upper (lower) panels correspond
    to solutions within the LMA (LMA-D) subset of parameter space.
    Notice that for marginalized $\eta$ no determination of the
    individual diagonal NSI terms beyond the $1.7\sigma$ level is
    possible, see text for details.}
  \label{fig:chisq-rng}
\end{figure}

We finish by quantifying the results of our analysis in terms of the
effective NSI parameters which describe the generalized Earth matter
potential and are, therefore, the relevant quantities for the study of
long-baseline experiments.  The results are shown in
Fig.~\ref{fig:chisq-rng} where we plot the dependence of the global
$\chi^2$ on each NSI effective couplings after marginalization over
all other parameters.\footnote{Notice that the correlations among the
  allowed values for these parameters are important and they are
  required for reconstruction of the allowed potential at given CL.}
Let us point out that, if only the results from Earth-based
experiments such as atmospheric, long-baseline and reactor data were
included in the analysis, the curves would be independent of
$\eta$. However, when solar experiments and COHERENT data are also
considered the global $\chi^2$ becomes sensitive to the value of
$\eta$.  Given that, what we quantify in Fig.~\ref{fig:chisq-rng} is
our present knowledge of the matter potential for neutrino propagation
in the Earth for \emph{any unknown value} of $\eta$. Technically this
is obtained by marginalizing the results of the global $\chi^2$ with
respect to $\eta$ as well, so that the $\Delta\chi^2$ functions
plotted in the figure are defined with respect to the absolute minimum
for any $\eta$ (which, as discussed above and shown in
Fig.~\ref{fig:chisq-eta}, lies close to $\eta \sim -45^\circ$).  In
the upper panels the oscillation parameters have been marginalized
within the LMA solution and in the lower ones within the LMA-D
solution.  Comparing the blue and cyan lines we conclude that COHERENT
has a sizable impact on the results for the LMA-D solution, whereas
within the LMA region its present contribution to the determination of
the generalized Earth matter potential is marginal.
Notice that, although in principle COHERENT allows to measure the
individual diagonal NSI (as seen in Fig.~\ref{fig:coh-range}) instead
of just their differences, such ability is lost for $\eta \approx
-35.4^\circ$, a value which is disfavored with respect to the global
best-fit point by $\Delta\chi^2 \simeq 3.0$ (see the cyan line in the
left panel of Fig.~\ref{fig:chisq-eta}). This implies that when $\eta$
is marginalized we always have
$\Delta\chi^2(\Eps_{\alpha\alpha}^\oplus) \leq 3$, so that no
determination of the diagonal NSI terms $\Eps_{\alpha\alpha}^\oplus$
is possible beyond the $1.7\sigma$ level.

Let us add that very recently the COHERENT collaboration has released
the full energy and time information of their event
rates~\cite{Akimov:2018vzs}, which we are not taking into account
here.  Concerning the energy spectrum, under the four fermion
interaction approximation (which we assume to hold at COHERENT) the
presence of NSI only induces an energy independent rescaling of the SM
prediction, so that including the energy information has no impact on
our results. As for the timing information, in principle it allows to
separate prompt and delay events (see for example
Ref.~\cite{Denton:2018xmq}) and provides therefore an extra handle on
the flavor of the NSI interactions. However, within the present
statistics of the experiment we expect such improvement to be
relatively modest, and to become even further diluted once combined
with the oscillation data in the full NSI parameter space. In view of
this, the global bounds derived here should be regarded as somewhat
conservative in what respects to the status of the LMA-D solution,
whereas they should be rather robust for what concerns the preferred
LMA solution.

\section{Summary}
\label{sec:summary}

In this work we have presented an updated analysis of neutrino
oscillation results with the aim of establishing how well we can
presently determine the size and flavor structure of NSI-NC which
affect the evolution of neutrinos in a matter background. In
particular we have extended previous studies by considering NSI with
an arbitrary ratio of couplings to up and down quarks (parametrized by
an angle $\eta$) and a lepton-flavor structure independent of the
quark type (parametrized by a matrix $\Eps_{\alpha\beta}^\eta$). We
have included in our fit all the solar, atmospheric, reactor and
accelerator data commonly used for the standard $3\nu$ oscillation
analysis, with the only exception of T2K and NO$\nu$A appearance data
whose recent hints in favor of CP violation are not easily
accommodated within the CP-conserving approximation assumed in this
work. In addition, we have considered the recent results on coherent
neutrino--nucleus scattering from the COHERENT experiment.  We have
found that:
\begin{itemize}
\item classes of experiments which are sensitive to NSI only through
  matter characterized by a limited range of proton/neutron ratio
  $Y_n$ unavoidably exhibit suppression of NSI effects for specific
  values of $\eta$. This is the case for solar data at $-70^\circ
  \lesssim \eta \lesssim -60^\circ$, for Earth-based (atmospheric,
  long-baseline, reactor) experiments at $\eta \approx -44^\circ$, and
  for COHERENT scattering data $\eta \approx -35^\circ$. Such
  cancellations limit the sensitivity to the NSI couplings;

\item moreover, the interplay between vacuum and matter contributions
  to the flavor transition probabilities in classes of experiments
  with limited energy range and/or sensitive only to a specific
  oscillation channel spoils the accurate determination of the
  oscillation parameters achieved in the standard $3\nu$
  scenario. This is particularly visible in $\Dmq_{21}$ and
  $\theta_{12}$ as determined by solar and KamLAND data, as well as in
  $\Dmq_{31}$ and $\theta_{23}$ as determined by atmospheric, LBL-CPC
  and MBL reactor data;

\item however, both problems can be efficiently resolved by combining
  together different classes of experiments, so to ensure maximal
  variety of matter properties, energy ranges, and oscillation
  channels.  In particular, our calculations show that the precise
  determination of the vacuum parameters is fully recovered (except
  for $\theta_{12}$) in a joint analysis of solar and Earth-based
  oscillation experiments, even when arbitrary values of $\eta$ are
  considered;

\item the well-known LMA-D solution, which arises in the presence of
  of NSI as a consequence of CPT invariance, is allowed at $3\sigma$
  for $-38^\circ \lesssim \eta \lesssim 87^\circ$ from the global
  analysis of oscillation data. The inclusion of the COHERENT results
  considerably improves this situation, however even in that case the
  LMA-D region remains allowed at the $3\sigma$ level for $-38^\circ
  \lesssim \eta \lesssim 14^\circ$.
\end{itemize}
In addition, we have determined the allowed range of the NSI couplings
$\Eps_{\alpha\beta}^\eta$ as a function of the up-to-down coupling
$\eta$, showing that such constraints are generically robust except
for a few specific values of $\eta$ where cancellations
occurs. Finally, in view of the possible implications that generic
NSI-NC may have for future Earth-based facilities, we have recast the
results of our analysis in terms of the effective NSI parameters
$\Eps_{\alpha\beta}^\oplus$ which describe the generalized matter
potential in the Earth, and are therefore the relevant quantities for
the study of atmospheric and long-baseline experiments.

\section*{Acknowledgements}

This work is supported by USA-NSF grant PHY-1620628, by EU Networks
FP10 ITN ELUSIVES (H2020-MSCA-ITN-2015-674896) and INVISIBLES-PLUS
(H2020-MSCA-RISE-2015-690575), by MINECO grant FPA2016-76005-C2-1-P
and MINECO/FEDER-UE grants FPA2015-65929-P and FPA2016-78645-P, by
Maria de Maetzu program grant MDM-2014-0367 of ICCUB, and by the
``Severo Ochoa'' program grant SEV-2016-0597 of IFT.
I.E.\ acknowledges support from the FPU program fellowship
FPU15/03697.

\appendix

\section{Details of the IceCube fit}
\label{sec:icecube}

The number of events measured by the IceCube detector have been
provided in a grid with 210 bins~\cite{Jones:2015,
  TheIceCube:2016oqi}, which depends on the reconstructed neutrino
energy (logarithmically spaced in 10 bins ranging from 400~GeV to
20~TeV) and the reconstructed neutrino direction (divided into 21
bins, with the first one defined as $-1 \leq \cos\Theta_\nu \leq
-0.96$ and the other 20 linearly spaced from $\cos\Theta_\nu =-0.96 $
to $\cos\Theta_\nu = 0.24$). To reproduce the number of events of each
bin we have computed
\begin{equation}
  N_i[\phi^\text{atm}] = \sum_\pm \int dE_\nu\, d\cos\Theta_\nu \,
  \phi^\text{atm}_{\mu,\pm}(E_\nu,\Theta_\nu) \,
  \langle P_{\mu\mu}^\pm(E_\nu,\Theta_\nu) \rangle \,
  A^\text{eff}_{i,\pm}(E_\nu,\Theta_\nu)
\end{equation}
where $\phi_{\mu,\pm}^\text{atm}(E_\nu,\Theta_\nu)$ is the atmospheric
muon neutrino flux for neutrinos ($+$) and anti-neutrinos ($-$). Among
the different alternatives provided by the IceCube collaboration we
have chosen to consider those tagged as ``initial'', which do not
include propagation effects across the Earth. Here
$A^\text{eff}_{i,\pm}(E_\nu,\Theta_\nu)$ is the effective area
encoding the detector response to a $\nu_\mu$ with energy $E_\nu$ and
direction $\Theta_\nu$ for the bin `$i$'. As effective area we have
used the nominal detector response. The quantity $\langle
P_{\mu\mu}^\pm(E_\nu,\Theta_\nu)\rangle$ is the flavor oscillation
probability averaged over the altitude of the neutrino production
point, defined as:
\begin{equation}
  \langle P_{\mu\mu}^\pm(E_\nu,\Theta_\nu) \rangle
  = e^{-\sum_n X_n(\Theta_\nu) \, \sigma_n^\pm(E_\nu)} \int dh \,
  P_{\mu\mu}^\pm(E_\nu,\Theta_\nu, h) \,
  \kappa^{\pm}(E_\nu,\Theta_\nu,h)
  \label{eq:Pmm}
\end{equation}
where $\kappa^\pm(E_\nu,\Theta_\nu,h)$ is the altitude distribution of
the flux normalized to one~\cite{Honda:2015fha}, $X_n(\Theta_\nu)$ is
the column density along the neutrino trajectory for the nucleon $n
\in \lbrace \text{proton}, \text{neutron} \rbrace$ and
$\sigma_n^\pm(E_\nu)$ is the corresponding inclusive cross-section for
$\nu_\mu$. Hence $\langle P_{\mu\mu}^\pm(E_\nu,\Theta_\nu) \rangle$
also includes the neutrino flux absorption by the Earth.

In order to reproduce the published fit~\cite{TheIceCube:2016oqi} we
need to include in the $\chi^2$ the contribution the systematic
uncertainties for every point in the parameter space. Such systematics
are included by the collaboration either as a discrete or a continuous
nuisance parameter. In our analysis all the systematics are treated as
continuous quantities and their effects on the number of events are
assumed to be linear. We can divide systematics into two classes:
those related to the neutrino flux, and those related to the detector
response and the optical properties of the ice.  The atmospheric
neutrino flux uncertainties are
\begin{itemize}
\item the normalization ($N_0$) which we assume to be unconstrained;

\item the tilt of the energy spectrum, which is parametrized by
  including a factor $(E_\nu/E_0)^\gamma$ with $E_0 = 1~\text{TeV}$, a
  5\% error on the power law index $\gamma$ and a central value
  $\gamma = 0$;
  
\item the ratio between the pion and the kaon decays contribution to
  the flux ($R_{\pi/K}$) with a 10\% error;

\item the ratio between the neutrino and the anti-neutrino flux
  ($\phi_{\nu}/\phi_{\bar\nu}$) with a 5\% error.
\end{itemize}
The uncertainties associated with the detector response and the ice
properties, which are provided by the collaboration in data sets using
the same grid as the effective area, are:
\begin{itemize}
\item the efficiency of IceCube Digital Optical Modules, where as
  nominal value we have used the table corresponding to 99\%
  efficiency, and as $1\sigma$ deviation we have used the table
  corresponding to 95\% efficiency;

\item the photon scattering in the ice, where the $1\sigma$ deviation
  is defined from the table corresponding to a 10\% increase with
  respect to the nominal response;
  
\item the photon absorption in the ice, where the $1\sigma$ deviation
  is defined as a 10\% increase in the absorption rate with respect to
  the nominal response;
  
\item the azimuthal anisotropy in the scattering length due to the
  dust grain shear; here the $1\sigma$ deviation is obtained from the
  data set denoted `SPICELEA ice model';
  
\item the optical properties of the ice column surrounding each
  string, where the $1\sigma$ deviation is obtained from the data set
  labelled `SPICEMIE ice model' which does not include hole ice
  effects.
\end{itemize}
For each point in the parameter space the $\chi^2[\phi^\text{atm}]$
value corresponding to the assumed flux model is calculated from the
theoretical predictions and the experimental values by means of a
log-likelihood function. The final $\chi^2$ for such point is then
chosen by minimizing over all the seven flux models provided by the
IceCube collaboration.

\section{Addendum: impact of new data (until July 2020)}

\begin{figure}\centering
  \includegraphics[width=0.8\textwidth]{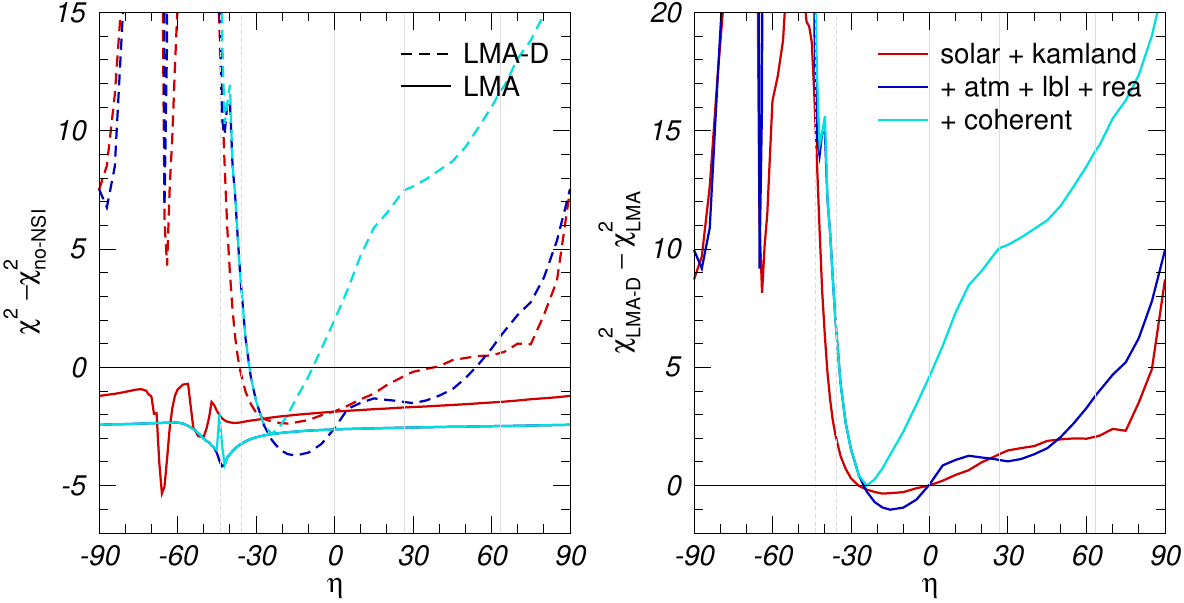}
  \caption{Left: $\chi^2_\text{LMA}(\eta) - \chi^2_\text{no-NSI}$
    (full lines) and $\chi^2_\text{LMA-D}(\eta) -
    \chi^2_\text{no-NSI}$ (dashed lines) for the analysis of different
    data combinations (as labeled in the figure) as a function of the
    NSI quark coupling parameter $\eta$. The full dark blue and light
    blue curves lie on top of each other.  Right:
    $\chi^2_\text{dark}-\chi^2_\text{light}\equiv
    \chi^2_\text{LMA-D}(\eta) - \chi^2_\text{LMA}(\eta)$ as a function
    of $\eta$. See text for details.}
  \label{fig2:chisq-eta}
\end{figure}

In this addendum we re-assess the constraints on Non-Standard
Interactions (NSI) from the global analysis of neutrino oscillation
data after including the new results released since the publication of
this work~\cite{Esteban:2018ppq}, in particular those presented at the
Neutrino2020 conference.  The new data considered here includes the
total energy spectrum and the day-night asymmetry of the 2970-day SK4
solar neutrino sample~\cite{SK:nu2020}, as well as the latest results
from long-baseline (LBL) experiments T2K~\cite{Abe:2019vii,
  T2K:nu2020} and NOvA~\cite{Acero:2019ksn, NOvA:nu2020}. In addition,
we have updated the reactor experiments
Double-Chooz~\cite{DoubleChooz:2019qbj, DoubleC:nu2020} to 1276/587
days of far/near detector data and RENO~\cite{Bak:2018ydk,
  RENO:nu2020} to 2908 days of exposure.

\begin{pagefigure}\centering
  \includegraphics[width=0.85\textwidth]{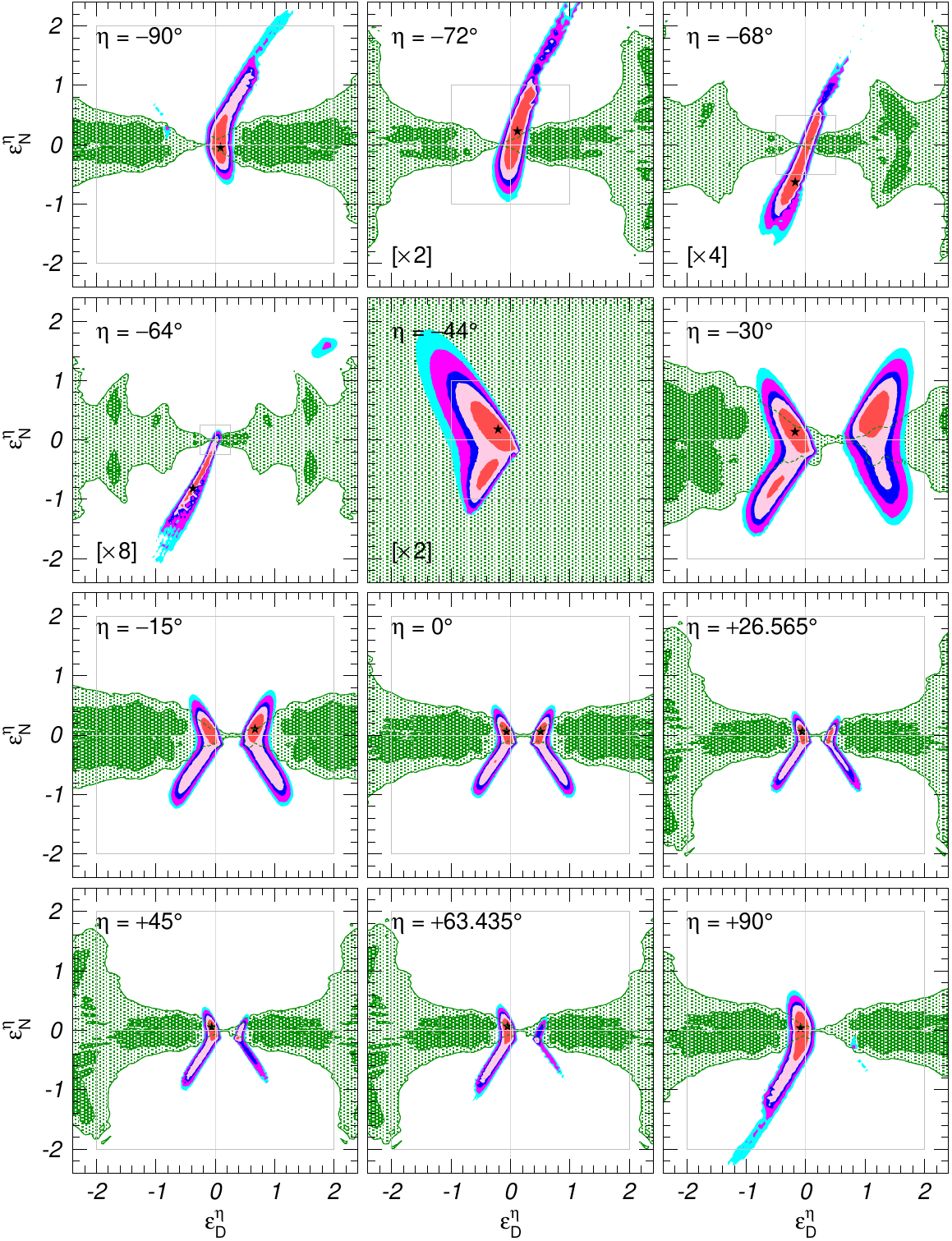}
  \caption{Two-dimensional projections of the $1\sigma$, 90\%,
    $2\sigma$, 99\% and $3\sigma$ CL (2~dof) allowed regions from the
    analysis of solar and KamLAND data in the presence of non-standard
    matter potential for the matter potential parameters
    $(\Eps_D^\eta, \Eps_N^\eta)$, for $\sin^2\theta_{13} = 0.022$ and
    after marginalizing over the oscillation parameters.  The best fit
    point is marked with a star. The results are shown for fixed
    values of the NSI quark coupling parameter $\eta$.  The panels
    with a scale factor ``$[\times N]$'' in their lower-left corner
    have been ``zoomed-out'' by such factor with respect to the
    standard axis ranges, hence the grey square drawn in each panel
    always corresponds to $\max\big( |\Eps_D^\eta|, |\Eps_N^\eta|
    \big) = 2$ and has the same size in all the panels.  For
    illustration we also show as shaded green areas the 90\% and
    $3\sigma$ CL allowed regions from the analysis of the atmospheric
    and LBL data. Note that, as a consequence of the periodicity of
    $\eta$, the regions in the first ($\eta = -90^\circ$) and last
    ($\eta = +90^\circ$) panels are identical up to an overall sign
    flip.}
  \label{fig2:sun-epses}
\end{pagefigure}

The main effect driven by the new results concerns the analysis of
solar and KamLAND data discussed in Sec.~\ref{sec:solar}.
As explained there, at the time of publication there was a tension of
$\Delta\chi^2\sim 7.4$ between these two data sets within the context
of the $3\nu$ oscillation analysis, arising from a combination of two
effects: (a) the $^8$B measurements performed by SNO, SK and Borexino
did not show any evidence of the low energy spectrum turn-up expected
in the standard LMA-MSW~\cite{Wolfenstein:1977ue, Mikheev:1986gs}
solution for the value of $\Dmq_{21}$ favored by KamLAND, and (b) the
observation of a non-vanishing day-night asymmetry in SK, whose size
was considerably larger than what predicted for the $\Dmq_{21}$ value
indicated by KamLAND.  Such tension could be alleviated in presence of
a non-standard matter potential, thus leading to a sizable decrease in
the minimum $\chi^2$ for the LMA solution for most values of $\eta$
($\Delta\chi^2\sim -7 \to -11$), as could be observed in the left
panel of Fig.~\ref{fig:chisq-eta}. Correspondingly, in
Fig.~\ref{fig:sun-epses}, which showed the two-dimensional projections
on the matter potential parameters ($\Eps_D^\eta$, $\Eps_N^\eta$) of
the $1\sigma$, 90\%, $2\sigma$, 99\% and $3\sigma$ CL (2~dof) allowed
regions from the analysis of solar and KamLAND data in the presence of
non-standard neutrino-matter interactions, the $3\nu$ standard LMA
oscillation scenario ($\Eps_D^\eta=\Eps_N^\eta=0$) was outside of such
allowed regions for most values of $\eta$.

As discussed in Ref.~\cite{Esteban:2020cvm}, with the updated SK4
solar data the tension between the best fit $\Dmq_{21}$ of KamLAND and
that of the solar results has decreased to $\Delta\chi^2_\text{solar}
= 1.3$. This is due to both the smaller day-night asymmetry, and the
slightly more pronounced turn-up in the low energy part of the
spectrum.  So now in the left panel in Fig.~\ref{fig2:chisq-eta} we see
that for the LMA solution the fit with NSI leads to a decrease of
about 1 unit of $\chi^2$ for most values of $\eta$.  Correspondingly
in Fig.~\ref{fig2:sun-epses} the $3\nu$ standard LMA oscillation
scenario, $\Eps_D^\eta=\Eps_N^\eta=0$ lies inside the $1\sigma$ LMA
allowed regions for most values of $\eta$.
Concerning the status of the LMA-D solution, the right panel in
Fig.~\ref{fig2:chisq-eta} shows that now LMA-D is allowed below
$3\sigma$ for $\eta>-40^\circ$ in the analysis of solar+KamLAND, for
$-38^\circ \lesssim \eta \lesssim 87^\circ$ in the global oscillation
analysis, and for $-38^\circ \lesssim \eta \lesssim 20^\circ$ when
including information from the total event rate at COHERENT.  From the
left panel we read that the best fit for the global analysis of
oscillations and also in combination with COHERENT corresponds to
$\eta\sim -45^\circ$ for LMA. For LMA-D the best fit for OSC (OSC+COH)
is obtained for $\eta\sim -15^\circ$ ($\eta\sim -20^\circ$).

\begin{figure}\centering
  \includegraphics[width=\textwidth]{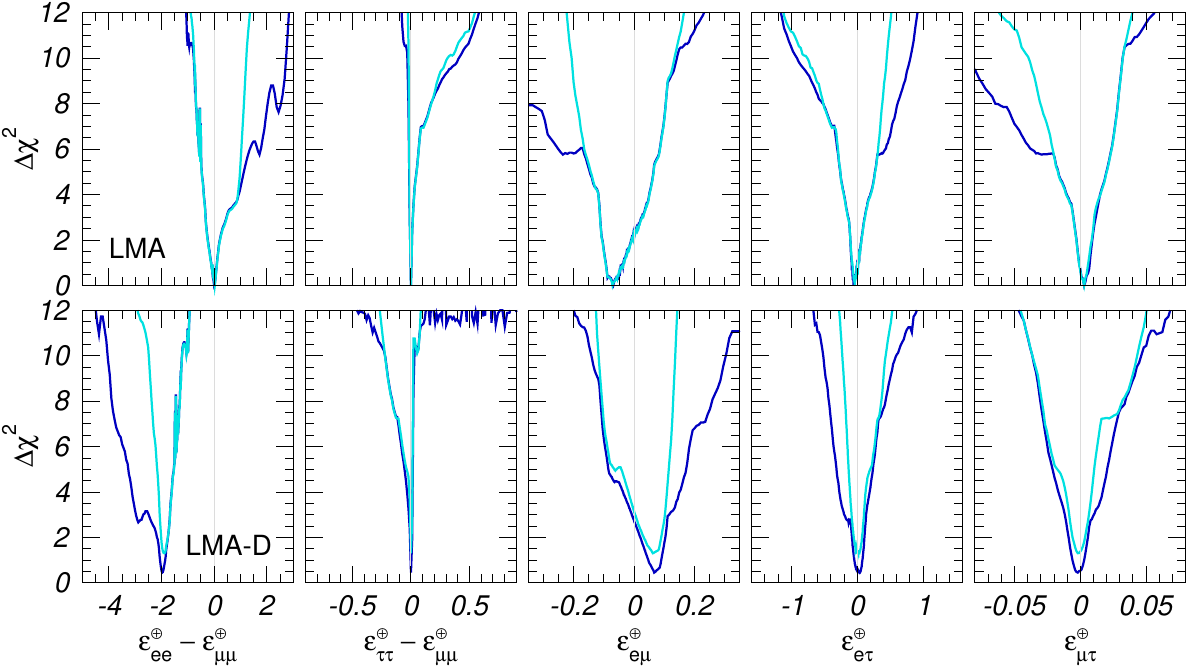}
  \caption{Dependence of the $\Delta\chi^2$ function on the effective
    NSI parameters relevant for matter effects in LBL experiments with
    arbitrary values of $\eta$, from the global analysis of solar,
    atmospheric, LBL-CPC and reactor data (blue lines) and including
    also COHERENT (cyan lines).  The upper (lower) panels correspond
    to solutions within the LMA (LMA-D) subset of parameter space.}
  \label{fig2:chisq-rng}
\end{figure}

In Fig.~\ref{fig2:chisq-rng} we plot the dependence of the global
$\chi^2$ on each NSI effective coupling relevant for neutrino
propagation in the Earth after marginalization over all other
parameters including $\eta$, so that the $\Delta\chi^2$ functions
plotted in the figure are defined with respect to the absolute minimum
for any $\eta$.  When compared with the corresponding results for the
old data shown in Fig.~\ref{fig:chisq-rng} we observe that, following
the discussion above, the minimum $\chi^2$ within LMA and LMA-D are
almost the same, while previously we had
$\Delta\chi^2_\text{min,LMA-D}\sim 3$. The other observable difference
is that including COHERENT has now a larger impact on the allowed
ranges in LMA.

Finally, for the sake of convenience and comparison with previous
results we list in the first columns in Table~\ref{tab2:ranges} the
95\% CL ranges for NSI with up-quarks only, down-quarks only, and
protons.  Generically the allowed ranges with in LMA are slightly
reduced and, as expected, the allowed ranges for
$\Eps_{ee}-\Eps_{\mu\mu}$ are now more symmetric around zero.

\begin{table}\centering
  \definecolor{grey}{gray}{0.75}
  \newcommand{\grsep}{~\color{grey}\vrule}
  \begin{tabular}{|l@{\grsep}r@{\grsep}r|l@{\grsep}l|}
    \hline
    \multicolumn{3}{|c|}{OSC}
    & \multicolumn{2}{c|}{${} + \text{COHERENT}$}
    \\
    \hline
    & $\text{LMA}$\hfil~
    & $\text{LMA}\oplus\text{LMA-D}$\hfil~ &
    & \hfil$\text{LMA} = \text{LMA}\oplus\text{LMA-D}$
    \\
    \hline
    \begin{tabular}{@{}l@{}}
      $\Eps_{ee}^u - \Eps_{\mu\mu}^u$ \\
      $\Eps_{\tau\tau}^u - \Eps_{\mu\mu}^u$
    \end{tabular}
    &
    \begin{tabular}{@{}r@{}}
      $[-0.072, +0.321]$ \\
      $[-0.001, +0.018]$
    \end{tabular}
    &
    \begin{tabular}{@{}r@{}}
      $\oplus [-1.042, -0.743]$ \\
      $[-0.016, +0.018]$       
    \end{tabular}
    &
    \begin{tabular}{@{}l@{}}
      $\Eps_{ee}^u$ \\
      $\Eps_{\mu\mu}^u$ \\
      $\Eps_{\tau\tau}^u$
    \end{tabular}
    &
    \begin{tabular}{@{}l@{}}
      $[-0.067, +0.547]$ \\
      $[-0.076, +0.455]$ \\
      $[-0.076, +0.455]$
    \end{tabular}
    \\
    $\Eps_{e\mu}^u$ & $[-0.050, +0.020]$ & $[-0.050, +0.059]$ &
    $\Eps_{e\mu}^u$ & $[-0.050, +0.020]$
    \\
    $\Eps_{e\tau}^u$ & $[-0.077, +0.098]$ & $[-0.111, +0.098]$ &
    $\Eps_{e\tau}^u$ & $[-0.077, +0.099]$
    \\
    $\Eps_{\mu\tau}^u$ & $[-0.006, +0.007]$ & $[-0.006, +0.007]$ &
    $\Eps_{\mu\tau}^u$ & $[-0.006, +0.007]$
    \\
    \hline
    \begin{tabular}{@{}l@{}}
      $\Eps_{ee}^d - \Eps_{\mu\mu}^d$ \\
      $\Eps_{\tau\tau}^d - \Eps_{\mu\mu}^d$
    \end{tabular}
    &
    \begin{tabular}{@{}r@{}}
      $[-0.084, +0.326]$ \\
      $[-0.001, +0.018]$
    \end{tabular}
    &
    \begin{tabular}{@{}r@{}}
      $\oplus [-1.081, -1.026]$ \\
      $[-0.001, +0.018]$
    \end{tabular}
    &
    \begin{tabular}{@{}l@{}}
      $\Eps_{ee}^d$ \\
      $\Eps_{\mu\mu}^d$ \\
      $\Eps_{\tau\tau}^d$
    \end{tabular}
    &
    \begin{tabular}{@{}l@{}}
      $[-0.063, +0.503]$ \\
      $[-0.072, +0.408]$ \\
      $[-0.072, +0.407]$
    \end{tabular}
    \\
    $\Eps_{e\mu}^d$ & $[-0.051, +0.020]$ & $[-0.051, +0.038]$ &
    $\Eps_{e\mu}^d$ & $[-0.050, +0.020]$
    \\
    $\Eps_{e\tau}^d$ & $[-0.077, +0.098]$ & $[-0.077, -0.098]$ &
    $\Eps_{e\tau}^d$ & $[-0.078, +0.098]$
    \\
    $\Eps_{\mu\tau}^d$ & $[-0.006, +0.007]$ & $[-0.006, +0.007]$ &
    $\Eps_{\mu\tau}^d$ & $[-0.006, +0.007]$
    \\
    \hline
    \begin{tabular}{@{}l@{}}
      $\Eps_{ee}^p - \Eps_{\mu\mu}^p$ \\
      $\Eps_{\tau\tau}^p - \Eps_{\mu\mu}^p$
    \end{tabular}
    &
    \begin{tabular}{@{}r@{}}
      $[-0.190, +0.927]$ \\
      $[-0.001, +0.053]$
    \end{tabular}
    &
    \begin{tabular}{@{}r@{}}
      $\oplus [-2.927, -1.814]$ \\
      $[-0.052, +0.053]$
    \end{tabular}
    &
    \begin{tabular}{@{}l@{}}
      $\Eps_{ee}^p$ \\
      $\Eps_{\mu\mu}^p$ \\
      $\Eps_{\tau\tau}^p$
    \end{tabular}
    &
    \begin{tabular}{@{}l@{}}
      $[-0.222, +1.801]$ \\
      $[-0.248, +0.282] \oplus [+0.625, +1.551]$ \\
      $[-0.248, +0.281] \oplus [+0.646, +1.548]$
    \end{tabular}
    \\
    $\Eps_{e\mu}^p$ & $[-0.145, +0.058]$ & $[-0.145, +0.145]$ &
    $\Eps_{e\mu}^p$ & $[-0.145, +0.058]$
    \\
    $\Eps_{e\tau}^p$ & $[-0.238, +0.292]$ & $[-0.292, +0.292]$ &
    $\Eps_{e\tau}^p$ & $[-0.239, +0.293]$
    \\
    $\Eps_{\mu\tau}^p$ & $[-0.019, +0.021]$ & $[-0.021, +0.021]$ &
    $\Eps_{\mu\tau}^p$ & $[-0.019, +0.021]$
    \\
    \hline
  \end{tabular}
  \caption{$2\sigma$ allowed ranges for the NSI couplings
    $\Eps_{\alpha\beta}^u$, $\Eps_{\alpha\beta}^d$ and
    $\Eps_{\alpha\beta}^p$ as obtained from the global analysis of
    oscillation data (left column) and also including COHERENT
    constraints.  The results are obtained after marginalizing over
    oscillation and the other matter potential parameters either
    within the LMA only and within both LMA and LMA-D subspaces
    respectively (this second case is denoted as $\text{LMA} \oplus
    \text{LMA-D}$). Notice that once COHERENT data are included the
    two columns become identical in all cases since for NSI couplings
    to $f=u,d,p$ the LMA-D solution is only allowed above 95\% CL.}
  \label{tab2:ranges}
\end{table}

\bibliographystyle{JHEPmod}
\bibliography{references}

\end{document}